%% file: twophoton-ii.tex
\documentclass[reprint, aps, pra, twoside, floatfix, longbibliography, superscriptaddress]{revtex4-2}

\usepackage{amsmath, amssymb}
\usepackage{graphicx}

\usepackage{multirow}
\usepackage{simpler-wick}

\usepackage{hyperref}
\IfFileExists{hidelinks}{
	\hypersetup{hidelinks}
}{
	\hypersetup{colorlinks=true, allcolors=[rgb]{0,0,.4}}
}

\usepackage{orcidlink}

\input{notations}
\input{highlights}

\begin{document}

\title{Two-photon coupling via Josephson element. II.
Interaction dressing, \mbox{cross-Kerr coupling}, and limits of low-energy bosonic model}

\author{E.~V.~Stolyarov\,\orcidlink{0000-0001-7531-5953}}
\email[E-mail: ]{eugene.stolyarov@bitp.kyiv.ua}
\affiliation{Quantum Optics and Quantum Information Group, Bogolyubov Institute for Theoretical Physics, National Academy of Sciences of Ukraine, vul.\ Metrolohichna 14-b, Kyiv 03143, Ukraine}
\author{V.~L.~Andriichuk\,\orcidlink{0000-0001-6004-7175}}
\affiliation{Institute of Physics of the National Academy of Sciences, pr.\ Nauky 46, Kyiv 03028, Ukraine}
\author{A.~M.~Sokolov\,\orcidlink{0000-0002-2691-317X}}
\email[E-mail: ]{andriy145@gmail.com}
\affiliation{Institute of Physics of the National Academy of Sciences, pr.\ Nauky 46, Kyiv 03028, Ukraine}

\begin{abstract}
We study the interactions mediated by a symmetric superconducting quantum interference device~(SQUID), their renormalizations, and the applicability of the anharmonic oscillator (bosonic) model for a coupled phase qubit.
The latter dwells in its metastable well holding a number of anharmonic energy states.
The coupling SQUID can switch between the single- and two-photon interactions \emph{in situ}.
We find that, in the \emph{bosonic} two-photon regime, the cross-Kerr coupling never vanishes as it dresses due to asymmetry in the qubit potential and nonlinearity of the coupler.
Our quantitative results also depend on the bosonic approximation.
We approach determining its limits by finding the minimum number of coherent energy states required for a dressing.
For that, we lay out diagrams of the dressing virtual processes that climb the qubit ladder as high as possible.
Near the two-photon resonance with a coupled resonator, we systematically calculate other relevant renormalizations due to nonresonant interactions.
We provide verifiable predictions for the coupling rates.
Modified systems can be applied for two-photon detection and for quantum-nondemolition readout of a qubit with an asymmetrical potential.
\end{abstract}

\maketitle

\section{Introduction} \label{secIntroduction}

In a curious prediction of quantum theory, a real transition may occur in two steps with an intermediate state never occupied considerably;
the two legs of such a process are called \emph{virtual} transitions.
In fact, one can even describe the simplest excitation exchange due to linear interaction as in the Jaynes-Cummings model~\cite{jc1963} in terms of two virtual processes: one destroying a quantum in the first system, and the other creating it in another system.
In the respective approximation, the combined system is never found in the intermediate state.
In a less trivial example, two nonresonant exchange interactions with an auxiliary system mediate a real excitation exchange between two other systems~\cite{yan2018tunable};
a modification~\cite{sete2021floating,marxer2023long} of such a coupler powers a commercial quantum computer~\cite{abdurakhimov2024iqmgarnet}.

Virtual transitions can chain into more complex processes as well.
Combined with intrinsic nonlinearity and transitions to the higher excited levels, linear interactions between transmon qubits~\cite{koch2007transmon} give rise to $ZZ$ interaction~\cite{solgun2022zz}.
Essentially, $ZZ$ coupling is due to \emph{parametric processes}~\cite{boyd2008nonlinear} in which the system state does not change, but the system energy acquires dependency on the state of its parts.
In the context of cavity~\cite{haroche1999cavity,carbonaro1979canonical} and circuit quantum electrodynamics~\cite{blais2004cavity,blais2007quantum,blais2021circuit}, an analogous coupling is known as dispersive.
It arises due to virtual transition of a resonator photon into the qubit and back~\cite{blais2007quantum}.

Two-photon coupling~\cite{nakamura2001rabi,lisenfeld2007thesis,ashhab2006rabi,garziano2015multiphoton,valimaa2022multiphoton,ayyash2024driven,parti} allows an electromagnetic degree of freedom to give up two photons in order to produce a single excitation in another system.
It can emerge---perturbatively---due to intermediate virtual processes as predicted~\cite{ashhab2006rabi,linskens1996twophoton,garziano2015multiphoton,sokolov2020superconducting,sokolov2023thesis} and measured~\cite{linskens1996twophoton} for different systems.
This type of nonlinear coupling can already find some applications.
Two-photon interaction facilitates single-mode squeezing~\cite{qin2024quantum}.
It may be used for novel quantum information processing protocols~\cite{ayyash2024driven} and achieving photon-number resolution in microwave photodetection~\cite{sokolov2020superconducting,sokolov2023thesis,stolyarov2025detector}.
With a deep strong two-photon coupling, interaction-induced spectral collapse may be observed~\cite{felicetti2015spectral}.

To achieve high two-photon coupling rates, one can either ``simulate'' it using the special driving schemes~\cite{qin2024quantum}, or seek a strong nonlinear interaction.
In the preceding contribution~\cite{parti}, we argued that a symmetrical dc SQUID can mediate nonperturbative two-photon coupling of reasonable strength that arises directly from the Josephson nonlinearity.
Independently, Ayyash \textit{et al.}\ conveyed the same point in Ref.~\cite{ayyash2024driven} for an asymmetrical SQUID.
The latter is better suited for coupling circuits with zero static displacement of the superconducting condensate phase over them, such as resonator and transmon~\cite{koch2007transmon} qubit.
However, in both works~\cite{parti,ayyash2024driven}, only the bare two-photon coupling was calculated;
i.e., the two-photon transitions due to possible virtual processes were not accounted for.

Moreover, neither of the works considered the cross-Kerr coupling.
It arises with a parametric process, similarly to the $ZZ$ or dispersive coupling mentioned above.
Typically, the term is used for two systems each having many energy states.
With the cross-Kerr coupling, the combined energy acquires a contribution proportional to each of the system populations.
That can be detrimental for implementing two-qubit gates~\cite{fors2024comprehensive,chakraborty2025tunable,marxer2023long,marxer2025fidelity}, applications of optomechanical systems~\cite{massel2015ckoptmech}, and in traveling-wave parametric amplifiers~\cite{esposito2021perspective}.
In the absorbing detectors of microwave photons~\cite{chen2011microwave,govia2012theory,opremcak2018measurement,opremcak2020high,shnyrkov2023rfsquid,ilinskaya2024fluxqubit,oelsner2017detection,delia2023steppingcloseraxion,grimaudo2022jjaxion,inomata2016single,lescanne2020irreversible,balembois2024cyclically}, cross-Kerr coupling introduces a photon-number--dependent detuning that may complicate the absorption.
On the other hand, this very type of coupling is required for the quantum-nondemolition detection~\cite{guerlin2007progressive,johnson2010qnd,leek2010cavity,kono2018,essig2021multiplexedpnm,dassonneville2020number,curtis2021singleshotpnr,liu2020qndhybrid}, dispersive qubit readout~\cite{blais2004cavity}, and other applications for quantum computing with superconducting circuits~\cite{ciani2019hamiltoncompcpbcrosskerr,kim2026tunable}.
The cross-Kerr coupling can be used to simulate optomechanical coupling using an additional driven mode~\cite{shang2019nonreciprocity}.

It seems, that the cross-Kerr interaction can be switched off in a Josephson coupler as from Refs.~\cite{parti,stolyarov2025detector,khabipov2022noncentrosymmetric,vrajitoarea2020}.
The coupler mediates interactions between the ``coordinates'' $\varphi_\res$ and $\varphi_\atom$ via its potential energy, proportional to $\cos(\varphi_\res - \varphi_\atom + \delta)$, where $\delta$ is some equilibrium phase.
To achieve higher values of the bare two-photon interaction, one can set $\delta = \pi/2$ to maximize the $\varphi_\res^2 \varphi_\atom$ expansion term in the energy~\cite{parti,khabipov2022noncentrosymmetric}.
At the same time, one may expect that the cross-Kerr coupling is absent.
Indeed, the bare cross-Kerr coupling only arises from the even-order expansion terms, absent at $\delta = \pi/2$.

In this work, we continue to study a resonator and a phase qubit coupled via the SQUID coupler (see Fig.~\ref{figCircuit}).
Using the bosonic approximation, we devise a systematic theory to answer three questions:
i) Having a plethora of different interactions and nonlinearities in that system, do they produce virtual transitions that yield a considerable correction to the two-photon coupling?
ii) Analogously, are there sizeable corrections in the cross-Kerr coupling? Can one mitigate this coupling, and if not, how big is it?
iii) For the virtual processes in the corrections, how many energy levels are required in the qubit metastable well?

The paper is organized as follows.
We review the physics and the main parameters of the circuit in Sec.~\ref{secCircuit} and discuss important bare interactions in Sec.~\ref{secBareInteractions}.
Then, in Sec.~\ref{secEffectiveModel}, we present the effective two-photon Hamiltonian and outline the approximations we use.
In Sec.~\ref{secInteractions}, we elucidate how the two-photon and cross-Kerr interactions are renormalized by other couplings.
We provide some involved virtual processes in the form of diagrams.
In Sec.~\ref{secEstimates}, we estimate the coupling rates for a reasonable set of circuit parameters.
Section~\ref{secDiscussion} discusses our results and their applicability, including the bosonic approximation.
We assess some applications and discuss generalizations of the theory in Sec.~\ref{secApplications} and conclude in Sec.~\ref{secConclusions}.

We delegate technical details to appendices.
In Appendix~\ref{apCircuitHamiltonian}, we provide our starting Hamiltonians.
Derivation of the effective two-photon Hamiltonian that accounts for all relevant renormalizations is in Appendix~\ref{apRenormHamiltonianDerivation}.
Appendix~\ref{apWick} provides an alternative calculation of the main cross-Kerr correction using Wick's theorem.
We also provide further detail on laying out the diagrams for the virtual processes involved.
In Appendix~\ref{apDynamicalEqs}, to show that this correction also arises in the classical limit, we use equations of motion to obtain it.
Appendix~\ref{apRules} outlines the recipe for finding the processes that climb the energy ladder as high as possible.
Appendix~\ref{apOtherInteractions} provides expressions and interpretations for non-essential corrections to the interaction rates.

\section{Circuit}
\label{secCircuit}

\begin{figure}
\includegraphics{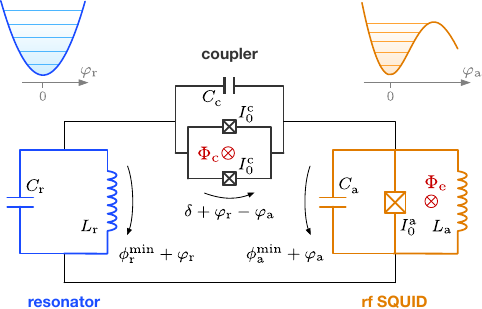}
\caption{A resonator and an rf SQUID in the regime of phase qubit interact through the SQUID coupler.
The system evolves around the equilibrium phase differences $\phi_\res^\mi$ and $\phi_\atom^\mi$.
The respective varying departures are $\varphi_\res$ and $\varphi_\atom$.
Conjugated to them, dynamic variables $n_\res$ and $n_\atom$ are the Cooper pair number at the full capacitances.
$\delta = \phi_\res^\mi - \phi_\atom^\mi$ is the equilibrium phase difference over the coupler.
$\Phi_\cpl$ and $\Phie$ are the coupler and rf SQUID bias fluxes.
Cartoons show potentials and energy levels of the resonator (left) and the phase qubit (right).
}
\label{figCircuit}
\end{figure}

It is convenient to describe the circuit dynamics in terms of dimensionless variables.
The number of Cooper pairs acts as a generalized momentum while the drop of the superconducting condensate phase acts as a generalized coordinate.

Consider the circuit in Fig.~\ref{figCircuit}.
For isolated resonator, $n_\res$ is the Cooper pair number on its capacitance $C_\res$.
Analogously, $n_\atom$ is the number of pairs on the rf SQUID full capacitance.
Near the equilibrium, only the departures from the static values of the condensate phases do matter.
$\varphi_\atom$ is such a departure of the phase difference across the rf SQUID junction.
$\varphi_\res$ is the phase departure across the resonator contacts.

We are interested in coupling between the resonator and the rf SQUID acting as an artificial atom.
The SQUID coupler mediates linear and nonlinear interactions.
Its Josephson energy
\begin{equation}
\label{eqCouplerJosephsonEnergy}
	E_\JJ^\cpl = 2E_{1\JJ}^\cpl \cos \frac{\pi\Phi_\cpl}{\Phi_0}
\end{equation}
can be tuned by the coupler bias $\Phi_\cpl$.
On the other hand, it is set by the Josephson energy of each of the coupler junctions, $E_{1\JJ}^\cpl = I_0^\cpl \Phi_0 / 2\pi$, where $I_0^\cpl$ is the critical current of the junction and $\Phi_0 = h/2e$ is the flux quantum.
The Josephson coupling energy reads $-E_\JJ^\cpl \cos(\varphi_\res - \varphi_\atom + \delta)$, where the static phase difference over the coupler
\begin{equation}
\label{eqCouplerPhase}
	\delta = \phi^\mi_\res - \phi^\mi_\atom - \frac{\pi\Phic}{\Phi_0}
\end{equation}
depends on its bias flux $\Phic$.
We denote the resonator and atom equilibrium phases by $\phi_\res^\mi$ and $\phi_\atom^\mi$.
The equilibrium is determined by the conservation of static currents as expressed in Appendix~\ref{apEquilibrium}.
Depending on the bias $\Phic$ and the equilibrium position, the Josephson coupling produces inductive linear interaction as well as various nonlinear processes, including the two-photon one~\cite{parti}.

In addition, the coupler charging mediates linear capacitive interaction as in Table~\ref{tablBareInteractions}.
Its charging energy is
\begin{equation}
\label{eqCouplerCapacitiveEnergy}
	E^\cpl_C = 4e^2 \frac{\tilde C_\cpl}{C'_\atom C_\res},
\end{equation}
where the coupler capacitance $C'_\cpl$ is loaded by the series connection with the resonator and atom capacitances, $\tilde C_\cpl = (1/C'_\cpl + 1/C_\res + 1/C'_\atom)^{-1} \approx C'_\cpl$.
The latter approximation qualifies as we have $C'_\cpl \ll C_\res, C'_\atom$ in Sec.~\ref{secEstimates}, similarly to the preceding work~\cite{parti}.
The full coupling capacitance $C'_\cpl = C_\cpl + 2C_{1\JJ}^\cpl$ consists of the shunt $C_\cpl$ and two junction capacitances $C_{1\JJ}^\cpl$ of the coupler.

We consider charging of the circuit capacitances to accumulate the ``kinetic'' energy of the circuit.
We denote $E_C^\res = e^2 / 2\tilde C_\res$ the resonator charging energy, where $\tilde C_\res = C_\res + C'_\atom C'_\cpl / (C'_\atom + C'_\cpl)$ is the loaded resonator capacitance.
That is, $C_\res$ is loaded by the parallel connection with the atom full capacitance $C'_\atom$ which is connected in series with the full coupling capacitance $C'_\cpl$.
Analogously, for the artificial atom $E_C^\atom = e^2 / 2\tilde C_\atom$, where $\tilde C_\atom = C'_\atom + C_\res C'_\cpl / (C_\res + C'_\cpl)$ and $C'_\atom = C_\atom + 2C_{1\JJ}^\atom$ is the full atom capacitance.
It consists of the shunting capacitance $C_\atom$ and the internal capacitances  $C_{1\JJ}^\atom$ of each of the atom junctions.
We rely on the exact expression for $\tilde C_\atom$ to tune the atom and resonator into the two-photon resonance.

Quadratic potential energy of the circuit arises as from linear inductances.
The Josephson energy also contributes.
The full inductive-like energies read
\begin{subequations}
\label{eqInductiveEnergiesRenorm}
\begin{gather}
\label{eqRenormELr}
	\tilde E^\res_L = E^\res_L + E^\cpl_\JJ \cos\delta,
\\
\label{eqRenormELa}
	\tilde E^\atom_L = E^\atom_L + E^\atom_\JJ \cos\phi^\mi_\atom
						+ E^\cpl_\JJ \cos\delta.
\end{gather}
\end{subequations}
Here $E_\JJ^\atom = I_0^\atom \Phi_0 / 2\pi$ is the atom Josephson energy.
$E_L^\res = \Phi_0^2 / 4 \pi^2 L_\res$ is the resonator inductive energy, and a similar expression holds for the atom inductive energy.
According to Eq.~\eqref{eqRenormELr}, the resonator inductance $L_\res$ is loaded by the coupler Josephson inductance, $L_\cpl = \Phi_0^2 / (4\pi^2 E_\JJ^\cpl \cos\delta)$, as $\tilde L_\res^{-1} = L_\res^{-1} + L_\cpl^{-1}$.
Equation~\eqref{eqRenormELa} expresses both the load from the coupler and the analogous load from the atom Josephson energy.

The resonator and atom degrees of freedom oscillate around the equilibrium.
With the nonlinearities switched off, they oscillate with the plasma frequencies
\begin{equation}
\label{eqPlasmas}
	\omega^p_\res = \frac 1\hbar \sqrt{8 E_C^\res \tilde E_L^\res},
\quad
	\omega^p_\atom = \frac 1\hbar \sqrt{8 E_C^\atom \tilde E_L^\atom}.
\end{equation}
Here, the loaded inductive energies $\tilde{E}^\res_L$ and $\tilde{E}^\atom_L$ depend on the equilibrium point as in Eqs.~\eqref{eqInductiveEnergiesRenorm}.

\section{Bare interactions}
\label{secBareInteractions}

We expand the Josephson energy in the circuit to the fourth order in the resonator and atom phase variables.
We provide the expansion in Appendix~\ref{apCircuitHamiltonian} after circuit quantization.
Here, we focus on the resulting interactions important for this work.

For the expansion of the Josephson energies to be valid, we require that the inductive-type energies dominate over the charging energies:
\begin{equation}
\label{eqPhaseIsGoodQuantumNumber}
	E^\res_L \ggg E^\res_C,
\quad
	\tilde E^\atom_L \ggg E^\atom_C.
\end{equation}
That assures the phase particles of both systems to be ``heavy'' and to reside close to the equilibrium.

We express the system variables in terms of the bosonic ladder operators:
$n_\res = i n_{\res, \zpf} (a^\dag - a)$ and $\varphi_\res = \varphi_{\res, \zpf} (a^\dag + a)$ for the resonator variables;
while $n_\atom = i n_{\atom, \zpf} (b^\dag - b)$ and $\varphi_\atom = \varphi_{\atom, \zpf} (b^\dag + b)$ for the atom variables.
Here, the magnitudes of the resonator vacuum fluctuations read
\begin{equation}
\label{eqResZPF}
	n_{\res, \zpf} = (\tilde{E}^\res_L/32 E^\res_C)^{1/4},
\quad
	\varphi_{\res, \zpf} = (2E^\res_C/\tilde{E}^\res_L)^{1/4},
\end{equation}
and the magnitudes of the atom vacuum fluctuations are
\begin{equation}
\label{eqAtomZPF}
	n_{\atom, \zpf} = (\tilde{E}^\atom_L/32 E^\atom_C)^{1/4},
\quad
	\varphi_{\atom, \zpf} = (2E^\atom_C/\tilde{E}^\atom_L)^{1/4}.
\end{equation}
In Appendix~\ref{apSecondQuantizedHamiltonian}, we write out the resulting ``second-quantized'' Hamiltonian in the fourth-order approximation of the Josephson energy.

In Table~\ref{tablBareInteractions}, we present the interactions dressing the two-photon, cross-Kerr, and linear coupling.
The energy terms in the table yield processes with the given rates or their multiples.
We depict some processes with diagrams.
When reading from right to left and from left to right, each diagram shows two Hermitian-conjugated processes.
Later on, we use such diagrams to illustrate the dressing and the minimal number of energy levels it requires.

\begin{table*}
\caption{Important bare interactions in the fourth-order approximation of the Josephson energy.
In the diagrams, wavy lines stand for creation or annihilation of a resonator photon while straight lines denote the atom processes.
Arrow lengths indicate change in the system energy.
In the two-photon regime, an atom excitation requires twice as much energy as a photonic one.
Each diagram reads from left to right and vice versa, as indicated by the arrows at the start and end of the respective processes.
(Potential asymmetry)
in the atom brings processes with amplitudes of order $\varphi_{\atom,\zpf}$ in the atom zero-point fluctuations, as explained in the text.
(Linear coupling):
Combined, the capacitive and inductive couplings with rates $g_1^c$ and $g_1^i$ give rise to number-preserving and -nonpreserving exchange processes.
(Cubic coupling) gives rise to the cross-Kerr processes which are always resonant.
(Quadratic couplings):
Two-photon interaction arises from a quadratic coupling, along with the other ones.
}
\label{tablBareInteractions}
\input{interactions-tabular.tex}
\end{table*}

The cubic nonlinearity $-\hbar X_\atom (b^\dag + b)^3$ in the atom potential arises from expansions of both the atom $-E_\JJ^\atom \cos(\phi_\atom^\mi + \varphi_\atom)$ and the coupling Josephson energies.
As $E_\JJ^\atom \varphi_{\atom,\zpf}^2 = (E_\JJ^\atom / 2\tilde E_L^\atom) \hbar\omega^p_\atom$, the third-order atom term $-(1/3!)E_\JJ^\atom \varphi_{\atom,\zpf}^3 \sin\phi_\atom^\mi$ turns first order in $\varphi_{\atom,\zpf}$ if $E_\JJ^\atom \gtrsim \tilde E_L^\atom$.
In that case, we have
\begin{equation}
\label{eqXaFull}
	X_\atom = \frac1{12} \frac{E^\atom_\JJ}{\tilde E_L^\atom}
				\omega^p_\atom \varphi_{\atom,\zpf} \sin\phi_\atom^\mi
		- \frac{E^\cpl_\JJ}{6\hbar} \varphi_{\atom,\zpf}^3 \sin\delta
\end{equation}
for the strength of the nonlinearity.

In Table~\ref{tablBareInteractions}, this cubic nonlinearity in the atom potential is characterized by
\begin{equation}
\label{eq:def_mu}
	\mu = \frac{12X_\atom}{\varphi_{\atom, \zpf} \, \omega^p_\atom}
		\approx \frac{\sin\phi_\atom^\mi}{\cos\phi_\atom^\mi + \beta_L^{-1}},
\end{equation}
where $\beta_L = E_\JJ^\atom / E_L^\atom$ is the screening parameter~\cite{clarkebraginsky2004squidhandbook} of the atom loop.
When using the expressions in the main part of the paper, to assess only the dominant corrections, one can neglect the coupler contribution as above.

Let us comment on some interactions in Table~\ref{tablBareInteractions}.
Linear coupling yields a number-nonpreserving single-photon interaction.
In it, both the atom and resonator simultaneously excite or de-excite by one energy level.
There is a similar two--atom-excitation process which does not preserve the energy.
Quadratic coupling also causes an optomechanical-type interaction~\cite{johansson2014optomechanical,eichler2018realizing}.
There, the atom plays a role of an optical cavity.
Its frequency is modulated by the generalized coordinate of the microwave resonator that plays a role of a mechanical mode.
In the ``reverse optomechanical'' interaction, the atom and resonator swap their roles as in Ref.~\cite{potts2025strong}.

Note the signs of some energy terms in Table~\ref{tablBareInteractions}.
We choose them so that the important interaction rates $g_2, K_0 > 0$ when the two-photon coupling strength $g_2$ attains its highest value (see Fig.~\ref{figCouplings}).
That is convenient in what follows, for reading whether the renormalization terms increase or decrease the rate magnitudes.
By the same reason, we choose $X_\atom > 0$ there, as it appears in the rate dressing.

\section{Effective model}
\label{secEffectiveModel}

In this section as well as in the next one, we do not detail the coupler regime.
That is, we provide the expressions for the coupling rates for an arbitrary coupler phase $\delta$ as given by Eq.~\eqref{eqCouplerPhase}.
We only require that one adjusts the artificial atom and the resonator to the two-photon resonance $\omega_\atom \approx 2\omega_\res$.
Here and below in the main text, $\omega_\atom$ and $\omega_\res$ stand for actual, dressed, frequencies of the atom and resonator.

\subsection{Two-photon Hamiltonian}
\label{secTwoPhotonHamiltonian}

In Appendix~\ref{apRenormHamiltonianDerivation}, we use Schrieffer-Wolff transformations to derive the system effective Hamiltonian close to the two-photon resonance,
\begin{equation}
\label{eqTwoPhotonHamiltonian}
\begin{split}
	\frac{\opH}{\hbar} = {}& \omega_\res a^\dag a
		+ \omega_\atom b^\dag b
		- \frac{\varXi_\atom}{2} b^{\dag 2} b^2
\\
		&{}- K_{0,X} a^\dag a b^\dag b - \tilde g_2 (a^{\dag2}b + b^\dag a^2).
\end{split}
\end{equation}
In the rest of this work, the two-photon coupling of rate $\tilde g_2$, the cross-Kerr coupling of rate $K_{0,X}$, and the atom anharmonicity $\varXi_\atom$ are most important for us.
Due to the two-photon coupling, annihilation of two resonator photons excites the artificial atom, and vice versa.
Both rates $\tilde g_2$ and $K_{0,X}$ are dressed by the other interactions, as we discuss in the next sections and in Appendix~\ref{apOtherInteractions}.
In contrast to the standard two-photon Jaynes-Cummings model~\cite{sukumar1981, *singh1982, vboas2019}, we describe the artificial atom as an oscillator of anharmonicity $\varXi_\atom$ rather than a two-level atom.

According to our preliminary calculations, there are also other, higher-order nonlinearities in Hamiltonian~\eqref{eqTwoPhotonHamiltonian} that can be important under reasonable conditions.
Namely, they are proportional to $\varXi_\atom^2 g_-^2 / (\omega_\atom - \omega_\res)^3$.
That is negligible in our case of interest with small anharmonicity $\varXi_\atom \ll \omega_\res, \omega_\atom$ and the linear coupling $g_-$ suppressed as in Sec.~\ref{secRateEstimates}.
However, these conditions do not hold in typical circuit quantum-electrodynamic setups with linear coupling as, e.g., in Refs.~\cite{jeffrey2014fast,barends2013xmon}.

Next, we outline the approximations in the derivation of the effective Hamiltonian.
Afterwards, in Sec.~\ref{secInteractions}, we approximate the two-photon and cross-Kerr interaction rates in the Hamiltonian~\eqref{eqTwoPhotonHamiltonian} and approximate the frequencies in Sec.~\ref{secResonatorParameters}.

\subsection{Approximations and the limits of validity}
\label{secApproximations}

As explained before, we expand the coupling terms in the Hamiltonian in the fourth order in the system phases.
In other words, we assume that the phase zero-point fluctuations are small, each system holds only a relatively small number of excitations, and the coupling energy is reasonably bounded from above.
We require that the inductive coupling is not in the ultrastrong regime, i.e., $g^i_1 \ll \omega^p_\atom$.
We rewrite that condition to obtain
\begin{equation}
\label{eqSizableEJcNoUltrastrong}
	\frac{E^\cpl_\JJ}{\tilde E^\atom_L} \ll \frac{2\varphi_{\atom,\zpf}}
												 {\varphi_{\res,\zpf}}
\end{equation}
using the expression for $g^i_1$ in Table~\ref{tablBareInteractions} with Eqs.~\eqref{eqPlasmas} and \eqref{eqAtomZPF}.

As already mentioned, we focus on the case of two-photon resonance, $\omega_\atom \approx 2\omega_\res$.
On the other hand, we assume that the atom single-photon transition is highly nonresonant and we treat the linear interaction as a perturbation.
Same applies to the optomechanical-type interactions in Table~\ref{tablBareInteractions}.
The system should not be driven in a way that induces transitions such as typically used for radiative cooling and amplification~\cite{aspelmeyer2014optomechanics,botter2012linear}.

We have mentioned for Eqs.~\eqref{eq:def_mu}, that we neglect the atom self-nonlinearity induced by the coupler while calculating the rate corrections.
Analogously, we neglect the resonator self-nonlinearity, as it is of purely induced character (see Appendix~\ref{apSecondQuantizedHamiltonian}).
As a result, we neglect small resonator anharmonicity in the Hamiltonian~\eqref{eqTwoPhotonHamiltonian} as discussed in Sec.~\ref{secDiscussionAnharmonicity}.
We assume that the induced nonlinearities are much weaker than the cubic nonlinearity in the atom potential.
Under the condition $E_\JJ^\cpl \ll \tilde E_L^\atom$, we still expect a reasonable accuracy in determining the coupling rates.
Also, we neglect the higher-order nonlinearities as explained in Sec.~\ref{secTwoPhotonHamiltonian}.

Finally, for calculations in Appendix~\ref{apRenormHamiltonianDerivation}, the artificial atom possesses an infinite number of the energy levels with the same anharmonicity.
An analogous approximation is routinely used for the transmon artificial atom~\cite{blais2021circuit}, with normally about ten energy levels~\cite{dumas2024ionization} within this model.
Recently, the limits of this approximation were studied for a transmon in dispersive readout~\cite{dumas2024ionization}.
In this paper, we make estimates for an rf SQUID with merely seven metastable states in Sec.~\ref{secEstimates}.
We further discuss the bosonic approximation in Sec.~\ref{secDiscussion}.

\section{Important interaction dressing}
\label{secInteractions}

Here we provide our analytical approximations for the resonant two-photon and cross-Kerr interactions as given in the Hamiltonian in Eq.~\eqref{eqTwoPhotonHamiltonian}, as well as for the nonresonant single-photon interaction.
With diagrams, we illustrate some virtual transitions that modify the bare interaction rates.
These diagrams are related to the Feynman diagrams, which were already studied in the context of Schrieffer-Wolff transformations~\cite{hillmann2022designingkerr,bravyi2011}.
Note that we do not show the real particles of an interaction with free legs as in Feynman diagrams.
Instead, we show the energy states the system virtually goes through during the full interaction process.
That allows us to clearly illustrate the smallest possible number of energy states for a given process to occur at all.

\subsection{Cross-Kerr coupling due to potential asymmetry}
\label{secCrossKerr}

We focus on an atom with asymmetric potential and nonlinear coupling.
In that case, the cross-Kerr coupling in the effective Hamiltonian~\eqref{eqTwoPhotonHamiltonian} is mainly corrected as
\begin{align}
\label{eqCrossKerrMainPartHalfBaked}
	K_{0,X} &= K_0 + \frac{24 g_2 X_\atom}{\omega^p_\atom}
\\
\label{eqCrossKerrMainPart}
	&= \frac{E^\cpl_\JJ}{\hbar}
		\big(\cos\delta - \mu \sin\delta \big)
		\varphi_{\res,\zpf}^2 \varphi^2_{\atom,\zpf}.
\end{align}
There is no correction from the reverse optomechanical processes of rate $G_2$.
That is the case as we neglect cubic anharmonicity in the resonator potential.

We provide the full derivation of Eq.~\eqref{eqCrossKerrMainPartHalfBaked} in Appendix~\ref{apSelfNonlinearitiesTreatment} using a Schrieffer-Wolff transformation as in Refs.~\cite{hillmann2022designingkerr, stolyarov2023pnr}.
A simpler derivation is sketched in Appendix~\ref{apWick}, where we rewrite the same transformation in a more convenient form.
Appendix~\ref{apDynamicalEqs} gives an even more straightforward calculation based on the Heisenberg equations of motion.
Following the calculation, the atom self-dressing, due to the cubic nonlinearity in its potential, spawns the cross-Kerr correction out of the optomechanical interaction of rate $g_2$.

\begin{figure*}[p]
\centering
\includegraphics{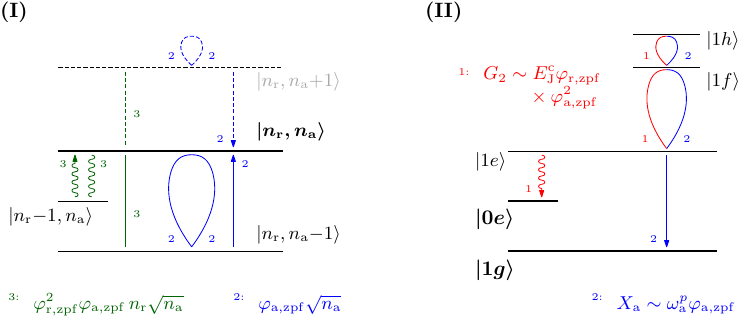}
\caption{Some corrections to the bare cross-Kerr and linear couplings, due to nonresonant bare processes in Table~\ref{tablBareInteractions}.
(I) Cross-Kerr correction due to optomechanical coupling $g_2$ with the energy quadratic in the resonator variables (3, green) combined with the cubic nonlinearity of the atom potential $X_\atom$ (2, blue).
Each part contributes to the full amplitude $\sim n_\res n_\atom $.
First index in a state label denotes the number of the resonator energy level while the second one labels the atom level.
The dashed lines correspond to an alternative process path that climbs the atom energy ladder as high as possible.
(II) Inductive single-photon correction due to combination of the interaction $G_2$ with the energy quadratic in the atom variables (1, red) and the atom cubic nonlinearity.
The full process is chosen to excite the atom as high as possible.
Loops contract creation and annihilation, such that only vacuum parts contribute---i.e., the unity in $bb^\dag = 1 + b^\dag b$.
The same-color loops do not rely on the availability of a next energy level.
Each virtual process brings a relevant factor (in corresponding colors and positions) to the full amplitude.
Other notations as in Table~\ref{tablBareInteractions}.
}
\label{figDiagramsLinearAndCrossKerr}
\end{figure*}

In a more physical picture, the correction in Eq.~\eqref{eqCrossKerrMainPartHalfBaked} occurs owing to the virtual processes.
The virtual optomechanical transitions of rate $g_2$ chain with the transitions caused by the atom nonlinearity.
We illustrate some example processes in Fig.~\ref{figDiagramsLinearAndCrossKerr}(I).

It is important here that two atom processes contract into a vacuum loop.
The operators from the same interaction type contract as in the Wick's theorem discussed in Appendix~\ref{apWick}.
In fact, such a single-colored loop only indicates the use of commutation relations to cast a bare interaction term or a Schrieffer-Wolff exponent into the normal-ordered form.
In the normal order, an interaction term uses the minimal number of the energy levels while still modeling the required coupling.

However, as discussed in Appendix~\ref{apWick}, we do not normal order the full combined virtual processes.
In the diagrams in this work, we present such processes that climb the atom energy ladder as high as possible.
In Appendix~\ref{apRules}, we give a detailed recipe for determining such processes.

\begin{figure*}[p]
\centering
\includegraphics{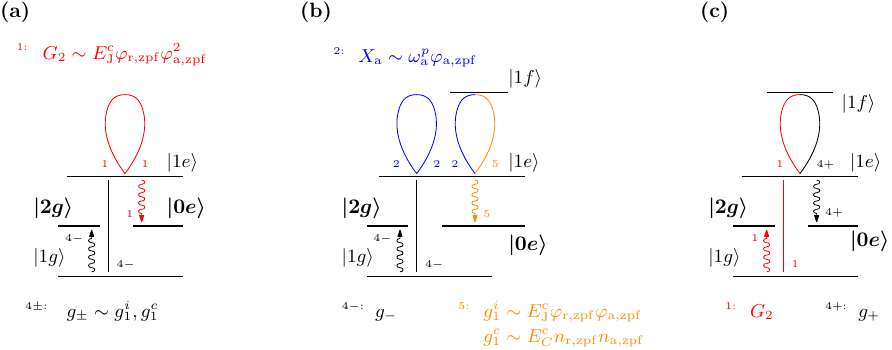}
\caption{
Some corrections to the two-photon coupling.
(a) Virtual transitions $0e \leftrightarrow 1e$ and $1e \leftrightarrow 2g$ mediated by the number-conserving linear coupling $g_-$ (4$-$ black) and the reverse optomechanical coupling $G_2$ (1, red).
(b) Virtual transitions $2g \leftrightarrow 1e$ with the number-conserving linear coupling as well as $1e \leftrightarrow 0e$ with the atom cubic nonlinearity $X_\atom$ (2, blue) and the number-nonconserving part of either $g_1^i$ the linear inductive or $g_1^c$ the linear capacitive coupling (5, orange).
(c) Virtual transitions $0e \leftrightarrow 1e$ and $1e \leftrightarrow 2g$ mediated by the two--atom-excitation coupling $G_2$ and the number-nonconserving linear inductive interaction $g_+$ (4$+$ black).
The depicted processes are chosen to excite the atom as high as possible.
Other notations as in Fig.~\ref{figDiagramsLinearAndCrossKerr} and Table~\ref{tablBareInteractions}.
}
\label{figDiagramsTwoPhoton}
\end{figure*}

We neglect other corrections in the cross-Kerr coupling.
They arise from nonresonant exchange interactions as for the dispersive shift in circuit quantum electrodynamics~\cite{blais2004cavity,blais2007quantum,blais2021circuit,gely2018nature,ansari2019superconducting}.
Dispersive corrections from the linear interaction are about $\varXi_\atom g_-^2 / (\omega_\atom - \omega_\res)^2$ or less.
That can be nonnegligible, but for $\omega_\atom \approx 2\omega_\res$ they only become comparable to the main cross-Kerr correction in Eqs.~\eqref{eqCrossKerrMainPartHalfBaked} and \eqref{eqCrossKerrMainPart} with high anharmonicities $|\varXi_\atom| / \omega_\res \gtrsim |g_2 X_\atom| / g_-^2 \approx \mu\omega_\atom \sin^2\delta / |12E_\JJ^\cpl \cos\delta|$.
This threshold value of $|\varXi_\atom| / \omega_\res$ is large near $\delta = \pi/2$ with large two-photon coupling $g_2$.
In addition, our preliminary calculations show, that for nonlinear coupling and asymmetry in the atom potential, there is another correction of about $\mu g_- G_2 \varphi_{\atom,\zpf} / \omega_\res$.
It can also matter where the two-photon coupling is large;
although, it becomes comparable to the main cross-Kerr correction when $|g_-| \gtrsim |X_\atom|$ for $\varphi_{\atom,\zpf} \sim \varphi_{\res,\zpf}$.
For the estimates in Sec.~\ref{secEstimates}, both dispersive corrections are small.

\subsection{Non-resonant linear interactions}
\label{secDipolarInteractions}

While we assume that the linear interactions are strongly nonresonant, they can become relevant if the coupler is used with a multimode resonator or a waveguide.
Here we discuss the dressing of the linear inductive interaction as mediated by the coupling SQUID.

According to Appendix~\ref{apTwoPhotonDerivation}, rate of the linear inductive interaction renormalizes as
\begin{align}
\label{eqInductiveCouplingRenormHalfBaked}
	\tilde g^i_1 &\approx g^i_1 - \frac{20 G_2 X_\atom}{\omega^p_\atom}
\\
\label{eqInductiveCouplingRenorm}
				 &= \frac{E^\cpl_\JJ}{\hbar}
						\Big(\cos\delta - \frac{5}{6}\mu \varphi^2_{\atom, \zpf} \sin\delta\Big)
	\varphi_{\res, \zpf} \varphi_{\atom, \zpf}.
\end{align}
We provide the dominant terms for both parts proportional to $\cos\delta$ and $\sin\delta$.
The first term in Eq.~\eqref{eqInductiveCouplingRenormHalfBaked} is the rate of bare inductive interaction due to the coupler Josephson inductance, as given in Table~\ref{tablBareInteractions}.
When $\delta \mod \pi = \pi/2$, this bare coupling vanishes.
Hence the second term in Eq.~\eqref{eqInductiveCouplingRenormHalfBaked} becomes dominant close to that point.
Equation~\eqref{eqInductiveCouplingRenorm} allows us to determine the coupling regime with vanishing single-photon interactions more precisely in Sec.~\ref{secRateEstimates}.

As follows from the calculation in Appendix~\ref{apSelfNonlinearitiesTreatment}, the second term in Eq.~\eqref{eqInductiveCouplingRenormHalfBaked} arises with the dressing of the atom in the harmonic approximation by the interactions with itself.
Namely, this occurs via its cubic nonlinearity as in Table~\ref{tablBareInteractions}.
That dressing spawns the correction to the linear coupling out of the interaction $G_2$ in the table.

The correction to the inductive single-photon coupling in Eqs.~\eqref{eqInductiveCouplingRenormHalfBaked} and \eqref{eqInductiveCouplingRenorm} can be understood further.
It arises as the virtual transitions due to the $G_2$ interaction chain with the transitions due to the atom nonlinearity.
We illustrate two such processes in Fig.~\ref{figDiagramsLinearAndCrossKerr}(II).

In the correction in Eq.~\eqref{eqInductiveCouplingRenorm}, four atom processes contract into two vacuum loops.
That allows for the required $\sqrt m$ coefficient when the full sequence acts on an atom excited state $\propto b^{\dag m} \ket g$.
Moreover, in Fig.~\ref{figDiagramsLinearAndCrossKerr}(II), we put the loops one above each other.
They correspond to the vacuum part of the $b^2 b^{\dag2}$ process when reading the diagram from right to left.
Note also how operators from different interactions can contribute to a vacuum loop.
In that case, additional energy level is required for the loop to form, as discussed in Appendix~\ref{apWick}.

In our model, there are no corrections to the capacitive single-photon coupling as in Table~\ref{tablBareInteractions}.
The reason is that both the atom nonlinearity and the quadratic couplings in Table~\ref{tablBareInteractions} are of the inductive type alone, i.e., they depend only on the atom phase variable $\propto b^\dag + b$.
When combining, they only result in inductive interactions.

\subsection{Two-photon coupling}
\label{secTwoPhotonIsBare}

We find that the corrections to the two-photon coupling are typically small.
We point out these corrections and argue that the two-photon coupling in the Hamiltonian~\eqref{eqTwoPhotonHamiltonian} can be approximated with its bare value as $\tilde g_2 \approx g_2$.

Recall that we consider bare interactions from the fourth-order expansion of the Josephson coupling energy.
The two-photon processes contain an odd number of creation and annihilation operators.
Therefore, a correction process involves either the $G_2$ process, or the $X_\atom$ atom self-nonlinearity process.
The lowest-order process with $G_2$ also requires a linear interaction as in Figs.~\ref{figDiagramsTwoPhoton}(a) and~(c).
The process with $X_\atom$ requires two linear interactions as in Fig.~\ref{figDiagramsTwoPhoton}(b). 
These process amplitudes are of the fifth order in the zero-point fluctuations.

However, these processes can be nonnegligible in our fourth-order expansion of the coupling.
The next-order expansion term adds $(g_2/2)(\varphi_{\res,\zpf}^2 + \varphi_{\atom,\zpf}^2)$ to the bare rate.
Consider the purely inductive corrections in Fig.~\ref{figDiagramsTwoPhoton} of about $(E_\JJ^\cpl/\hbar)^2 \varphi_{\res,\zpf}^2 \varphi_{\atom,\zpf}^3 / \omega_\res$, where the denominator takes into account nonresonant character of the process parts.
For a stronger Josephson coupling energy $E_\JJ^\cpl$, these corrections can exceed the neglected fifth-order expansion term linear in $E_\JJ^\cpl$.
That is the case when
\begin{equation} \label{eqNonnegligibleCorrections}
	4E_\JJ^\cpl \gg \hbar\omega_\res (1 + \varphi_{\res,\zpf}^2 \varphi_{\atom,\zpf}^{-2}).
\end{equation}
We explicitly provide these two-photon rate corrections and discuss them in detail in Appendix~\ref{apTwoPhotonFormula}.

Still, the two-photon corrections are small compared to the bare coupling $g_2$ under reasonable assumptions.
Amplitudes of the corrections in Figs.~\ref{figDiagramsTwoPhoton}(a) and~(c) are proportional to $g_\pm G_2 / \omega_\res$.
However, $g_\pm \ll \omega_\res$ as we require the coupling strength below the ultrastrong regime.
On the other hand, $G_2 \sim g_2$ according to Table~\ref{tablBareInteractions} once the resonator and the atom zero-point fluctuations are of the same order of magnitude, $\varphi_{\res,\zpf} \sim \varphi_{\atom,\zpf}$.
In the other case with $\varphi_{\res,\zpf} \ll \varphi_{\atom,\zpf}$, the corrections like in Figs.~\ref{figDiagramsTwoPhoton}(a) and (c) can be of the same order as the bare two-photon coupling.

The correction with $X_\atom$ as in Fig.~\ref{figDiagramsTwoPhoton}(b) arises even in the absence of nonlinear coupling.
However, it is even smaller than the other corrections, according to the expressions in Appendix~\ref{apTwoPhotonFormula}.
A similar process yields two-photon coupling with the rate $2\mu g_+ g_1^c \varphi_{\atom,\zpf} / 9\omega_\res$.
It involves capacitive coupling $g_1^c$ instead of the inductive $g_1^i$ in Fig.~\ref{figDiagramsTwoPhoton}(b).
In principle, it can be arbitrarily strong;
however, for arranging two-photon coupling it makes sense to limit $g_1^c \sim g_1^i$ as discussed in Appendix~\ref{apTwoPhotonFormula}.

\section{Numerical estimates}
\label{secEstimates}

In this section, we use our analytical results to calculate the coupling rates and their dressing.
As in the preceding part~\cite{parti}, we consider the artificial atom to dwell in its metastable well initially.
We tune the well to accommodate more than enough metastable states to sustain the dressing virtual transitions in Figs.~\ref{figDiagramsLinearAndCrossKerr} and \ref{figDiagramsTwoPhoton}.
At the same time, we seek the atom anharmonicity much stronger than the two-photon interaction rate.
This regime resembles the phase qubit in Ref.~\cite{bennett2009decoherencerfsquid}.

Also, we explain how we match the lowest two-photon resonance, i.e. for the $0e \leftrightarrow 2g$ transition.
We focus on the case of strong coupling, where the two-photon coupling rate $\tilde g_2$ exceeds the damping rates.
Hence, we consider the systems to be resonant when their detuning is below $\tilde g_2$, analogously to the results of Ref.~\cite{kosugi2005rabi} for the single-photon coupling.

\subsection{Phase qubit regime of the rf SQUID}
	\label{secAtomParameters}

\begin{figure}[t]
	\centering
	\includegraphics{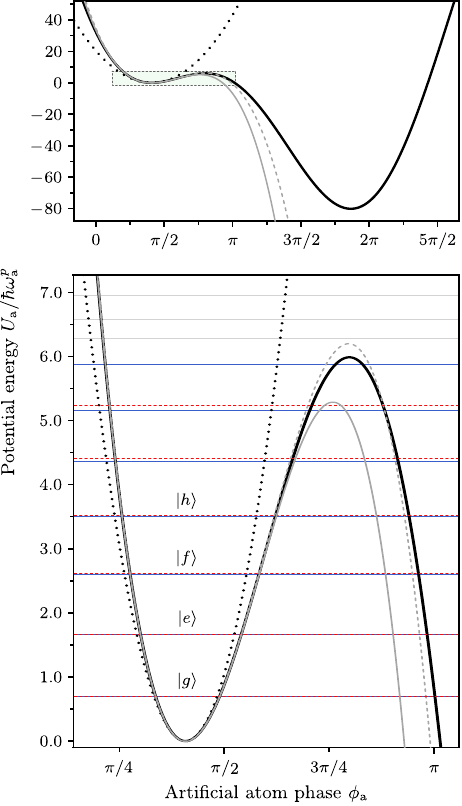}
	\caption{(Top panel) Double-well potential (thick black) of the uncoupled rf SQUID as given by Eq.~\eqref{eqAtomPotential}.
		Other curves show the potential in the harmonic (dotted), cubic (thin dashed), and quartic (thin solid) approximations.
		The energy origin is at the bottom of the left well.
		Shaded rectangle highlights the region shown in the other panel.
		(Bottom panel) Metastable well hosting levels $\ket g$, $\ket e$, $\ket f$, $\ket h$, and three others.
		Solid lines indicate the energies of metastable states and several over-the-barrier states, selected among numerical solutions of the stationary Schr\"odinger equation for the full potential.
		Dashed lines indicate the energy levels according to the perturbation theory as in Eq.~\eqref{eqFreqs}.
		Atom parameters: ratio of the capacitive and inductive energies $E^\atom_C/E^\atom_L = 3.950 \times 10^{-4}$, screening parameter $\beta_L = E^\atom_\JJ/E^\atom_L = 3.3424$, and external flux $\Phie = 0.7150\Phi_0$.
	}
	\label{figTwoWells}
\end{figure}

The exact expression for the isolated rf SQUID potential energy reads
\begin{equation}
\label{eqAtomPotential}
	U_\atom = \frac{E^\atom_L}{2}
		\left(\phi_\atom - \frac{2\pi\Phie}{\Phi_0}\right)^2
		- E^\atom_\JJ \cos\phi_\atom.
\end{equation}
The potential and energy levels are shown in Fig.~\ref{figTwoWells} for the uncoupled rf SQUID.
Turning on the Josephson coupling only renormalizes the atom Josephson inductance.
According to Eq.~\eqref{eqRenormELa}, that distorts the potential only slightly when $E_\JJ^\cpl \ll E_L^\atom + E_\JJ^\atom \cos\phi_\atom^\mi$.
For the energy ratios and external flux as in the Fig.~\ref{figTwoWells} caption, the atom equilibrium phase is $\phi_\atom^\mi \approx 0.4091 \pi$.
Hence the atom Josephson energy improves the criterion by about $E_\JJ^\atom \cos\phi_\atom^\mi \approx 0.9411 E_L^\atom$.

In this subsection and in Sec.~\ref{secResonatorParameters}, we first provide the energies and frequencies for the isolated systems.
That is, $E_\JJ^\cpl = 0$ both for determining the inductive energies in Eqs.~\eqref{eqInductiveEnergiesRenorm} and the equilibria in Appendix~\ref{apEquilibrium} while $C'_\cpl = 0$ for determining the capacitive energies.

We choose the rf SQUID inductance $L_\atom = 1.0000\,\mathrm{nH}$ ($E^\atom_L/h=163.46\,\mathrm{GHz}$) and critical current $I^\atom_0=1.1000\,\mathrm{\mu A}$ ($E^\atom_\JJ/h=546.35\,\mathrm{GHz}$) close to Ref.~\cite{opremcak2018measurement}.
However, we pick a smaller full capacitance of the artificial atom $C^\prime_\atom=300.0\,\mathrm{fF}$ ($E^\atom_C/h=64.57\,\mathrm{MHz}$) to increase its anharmonicity.
The rf SQUID external flux is indicated in the Fig.~\ref{figTwoWells} caption.
These parameters yield the plasma frequency of the metastable well $\omega^p_\atom = 2\pi\,12.80\,\mathrm{GHz}$ and the magnitudes of zero-point fluctuations $n_{\atom, \zpf} = 3.520$ and $\varphi_{\atom, \zpf} = 0.1420$.

Anharmonicity in the metastable well is $[(E_e - E_g) - (E_f - E_e)]/h = 391.28\,\mathrm{MHz}$, where $E_g$, $E_e$, and $E_f$ are the eigenenergies from numerical solution of the Schr\"odinger equation.
On the other hand, perturbative estimate with Eq.~\eqref{eq:AtomAnharm} yields $\varXi_\atom = 2\pi\,325.04\,\mathrm{MHz}$.
In the phase qubit regime, the main contribution to the anharmonicity arises from the cubic nonlinearity in the atom potential, as we detail below in Sec.~\ref{secResonatorParameters}.

Note, that for the capacitance density of $50\,\mathrm{fF/\mu m^2}$ in a Josephson junction and its critical current density of $1.7\,\mathrm{\mu A/\mu m^2}$ \cite{steff2006,lisenfeld2007thesis}, self-capacitance of the atom junction is about $32\,\mathrm{fF}$ for our choice of the critical current $I^\atom_0$.
Thus, to achieve the total capacitance of the rf SQUID as above, one needs a $170\,\mathrm{fF}$ shunting capacitance.

Notice how the perturbative energy levels in Fig.~\ref{figTwoWells} compare to the levels calculated numerically.
The perturbative results there are according to the second-quantized Hamiltonian~\eqref{eqAtomHamiltonian} in Appendix~\ref{apRenormHamiltonianDerivation}.
Discrepancies with the numerics are only visible close to the barrier top.
They occur as the potential nonlinearities become comparable to its harmonic approximation.
For the highest metastable level, the well becomes visibly wider than the harmonic approximation.
At the same time, the barrier becomes so thin that the wavefunctions should spread into it much deeper.
Hence the energies are below the perturbation theory result, as seen in the figure.

Moreover, the perturbative theory misses one metastable level altogether.
We calculate the perturbative energy levels for the fourth-order approximation to the potential.
It breaks vividly near the barrier top (see Fig.~\ref{figTwoWells}) and hence misses one metastable level.
Note that the cubic approximation happens to work better near the barrier top.
However, we use the fourth-order one, as we are more interested in predicting the lowest transition frequency analytically.

\subsection{Coupler parameters}
\label{secCouplerParameters}

We choose a moderate value of the coupler Josephson energy.
Each of its junctions with critical current $I^\cpl_0=30\,\mathrm{nA}$ possesses the energy of $E^\cpl_{1 \JJ}/h \approx 14.90\,\mathrm{GHz}$.
With capacitance $C^\prime_\cpl=5\,\mathrm{fF}$ ($E^\cpl_C/h \approx 2.70\,\mathrm{MHz}$), the coupler mediates a capacitive coupling with a rate of about $2\pi \, 50$\,MHz well-attainable in circuit quantum electrodynamics~\cite{jeffrey2014fast,zhao2020merged}.

We consider coupler Josephson energy that is relatively small compared to other respective energies that are relevant for determining the transition frequencies, $2E_{1\JJ}^\cpl \sim 0.1 (E_L^\atom + E_\JJ^\atom \cos\phi_\atom^\mi), 0.1 E_L^\res$.
Still, variation of this coupler energy may bring the system off-resonance.
Indeed, $2E_{1\JJ}^\cpl$ shifts both the atom and the resonator full inductive energies as in Eqs.~\eqref{eqInductiveEnergiesRenorm}.
Depending on the coupler bias $\Phi_\cpl$, that yields a variation in the detuning $\omega_\atom - 2\omega_\res \sim \omega^p_\atom - 2\omega^p_\res$ of up to about $E_{1\JJ}^\cpl |\omega^p_\atom/E_\JJ^\atom - 2\omega^p_\res/E_L^\res| \sim 0.1 \omega^p_\atom$.
Under normal circumstances, this far exceeds both the linewidths and the two-photon coupling rate $\tilde g_2$.
That said, we aim for the resonance to occur close to the coupler bias with the highest two-photon coupling rate $\tilde g_2$.

\subsection{Resonator parameters and analytics for the resonance position}
\label{secResonatorParameters}

We choose the resonator parameters to hit the resonance as described below.
In addition, we fix the resonator and atom zero-point phase fluctuations to have close amplitudes.
We arrive at the resonator inductance $L_\res = 0.84868\,\mathrm{nH}$ ($E^\res_L/h = 192.61\,\mathrm{GHz}$) and capacitance $C_\res = 692.8\,\mathrm{fF}$ ($E^\res_C/h = 27.05\,\mathrm{MHz}$).
For uncoupled resonator, this yields plasma frequency $\omega^p_\res = 2\pi \,6.56\,\mathrm{GHz}$ and zero-point fluctuations with amplitudes $n_{\res,\zpf} = 3.830$ and $\varphi_{\res,\zpf} = 0.1305$.

One may expect, that the two-photon coupling rate $\tilde g_2$ is close to its maximum when the inductive component of the coupling is fully turned on with $E_\JJ^\cpl = 2E^\cpl_{1\JJ}$.
According to Eq.~\eqref{eqCouplerJosephsonEnergy}, the latter occurs at zero coupler bias, $\Phic = 0$.
On the other hand, the main contribution to $\tilde g_2$ arises from its bare part $g_2$ which maximizes for $\delta \mod \pi = \pi/2$.
As the coupler static phase $\delta$ depends on $\Phic$ as in Eq.~\eqref{eqCouplerPhase}, these simple conditions do not necessarily coincide.
However, we rely on the rf SQUID static phase $\phi_\atom^\mi$ to provide the most of the required coupler phase $\delta$.
This assumption holds relatively well for the atom as in Sec.~\ref{secAtomParameters} (see also Fig.~\ref{figTwoWells}).
Hence, leaving $\Phi_\cpl = 0$ indeed yields reasonably high values of $\tilde g_2$ with such a value of $\phi_\atom^\mi$.

Therefore, we choose the resonator parameters above so that the two-photon resonance occurs for $\phi_\atom^\mi$ as in Sec.~\ref{secAtomParameters} and zero coupler bias, $\Phi_\cpl = 0$.
Additionally, with such a choice, our results are directly applicable for a single-junction coupler as discussed in Sec.~\ref{secRateEstimates}.

In the rest of this subsection, we verify the resonance condition with an analytical approximation.
We approximate the frequencies in the Hamiltonian~\eqref{eqTwoPhotonHamiltonian} as
\begin{equation}
\label{eqFreqs}
	\omega_\res \approx \omega^p_\res - \frac{K_{0,X}}2,
\quad
	\omega_\atom \approx \omega^p_\atom - \varXi_\atom
								- \frac{K_{0,X}}2.
\end{equation}
Here we omit other interaction-induced shifts that are of order of $g_\pm^2 / \omega_{\res,\atom}$ and $\varXi_\atom g_\pm^2 / \omega_{\res,\atom}^2$ or smaller.
We report in Sec.~\ref{secCrossKerr} that $K_{0,X}$ in Eqs.~\eqref{eqFreqs} is also the main contribution to the cross-Kerr coupling strength.
Another frequency shift is of the same magnitude as the atom anharmonicity
\begin{equation} \label{eq:AtomAnharm}
	\varXi_\atom = \frac{E^\atom_C}{\hbar}
		\Bigg[\frac53 \bigg(\frac{E^\atom_L}{\tilde{E}^\atom_L}\bigg)^2
						\Big(\phi^\mi_\atom - \frac{2\pi\Phie}{\Phi_0}\Big)^2
			  - \frac{E^\atom_L - \tilde E^\atom_L}{\tilde{E}^\atom_L} \Bigg].
\end{equation}
It arises from the same self-nonlinearities as the anharmonicity.
In the brackets in Eq.~\eqref{eq:AtomAnharm}, the first term comes from the cubic perturbation in the atom potential.
The second one is the quartic correction.
See Appendix~\ref{apRenormHamiltonianDerivation} for a derivation using the Schrieffer-Wolff transformations.
Note that the shift analogous to $\varXi_\atom$ is negligibly small in the resonator frequency $\omega_\res$ as we discuss in Sec.~\ref{secDiscussionAnharmonicity}.

According to Eqs.~\eqref{eqFreqs}, the resonator and the atom frequencies are $\omega_\res \approx 2\pi \, 6.74\,\mathrm{GHz} - K_{0,X}/2$ and $\omega_\atom \approx 2\pi \, 13.49\,\mathrm{GHz} - K_{0,X}/2$.
It is already clear that we hit the resonance when $K_{0,X}/2 < \tilde g_2$.
In fact, calculations yield $K_{0,X} \approx 2\pi \, 15.7\,\mathrm{MHz}$ and $|\omega_\atom - 2\omega_\res| \approx 2\pi \, 1.5\,\mathrm{MHz}$ while $\tilde{g}_2 \approx 2\pi \, 27.0\,\mathrm{MHz}$.

Note also that the quartic contribution to the anharmonicity in Eq.~\eqref{eq:AtomAnharm} provides a correction of order of magnitude of $E^\atom_C / \hbar$.
More precisely, it is about 30\,MHz with the artificial atom parameters as in Sec.~\ref{secAtomParameters}.
This correction exceeds $\tilde g_2$ and hence is important for us.
Same holds for the contribution of about 10\% in $\varXi_\atom$ due to the coupler Josephson inductance.

\subsection{Numerics for the resonance position}
\label{secResonance}

\begin{figure}
\includegraphics{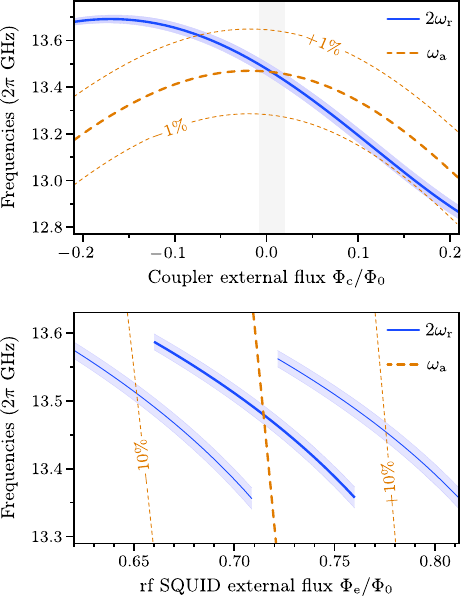}
\caption{Dependence of the resonator and atom frequencies on the flux biases: the coupler bias $\Phi_\cpl$ in the upper panel and the atom bias $\Phi_\ext$ in the lower panel.
Resonator frequency $\omega_\res$ according to Eq.~\eqref{eqFreqs};
atom frequency $\omega_\atom$ determined numerically as described in the text.
Blue area is the region from $2\omega_\res - \tilde g_2$ to $2\omega_\res + \tilde g_2$, where $\tilde g_2$ is the two-photon coupling strength.
Gray shade marks the minimal region of the $0e \leftrightarrow 2g$ resonance, i.e., where the detuning is smaller than $\tilde g_2$.
Circuit parameters as in the text.
Thin lines show the frequencies when the atom Josephson energy departs from the required value by the indicated fraction.
}
\label{figResonance}
\end{figure}

Furthermore, we study the resonance position numerically in Fig.~\ref{figResonance}.
It demonstrates how the system frequencies depend on the coupler bias $\Phi_\cpl$ and the atom bias $\Phi_\ext$.
Near the curve intersections, the two-photon transition $0e \leftrightarrow 2g$ is resonant.
To determine the atom frequency in the figure, we solve the Schr\"odinger equation for the rf SQUID potential~\cite{parti}
\begin{equation}
\label{eqAtomPotentialPerturbed}
	U'_\atom = U_\atom - E^\cpl_\JJ \cos\Big(\phi_\res^\mi - \phi_\atom - \frac{\pi\Phic}{\Phi_0}\Big)
\end{equation}
that is perturbed by the coupler.
Here the potential $U_\atom$ of the uncoupled atom is given by Eq.~\eqref{eqAtomPotential}.
For each value of the bias $\Phic$, we determine $\phi_\res^\mi$ and $\phi_\atom^\mi$ from the equations for equilibrium in Appendix~\ref{apEquilibrium}, before solving the Schr\"odinger equation.

The upper panel in Fig.~\ref{figResonance} supports our analytical estimates.
For $\Phi_\cpl = 0$, the atom frequency is $\omega_\atom \approx 2\pi \, 13.47\,\mathrm{GHz} - K_{0,X}/2$.
That differs from the analytical prediction by about $2\pi \, 10\,\mathrm{MHz}$.
Mostly, this error arises as the analytics underestimates the atom anharmonicity (see Sec.~\ref{secAtomParameters}).
We have verified that, for a smaller number of metastable energy levels, this discrepancy increases.

A more fundamental discrepancy occurs with the variation of the atom Josephson energy $E_\JJ^\atom$.
It is caused by fabrication imperfections, primarily fluctuations of the junction area and inhomogeneities in the insulating barrier~\cite{osman2021simplified, osman2023mitigation, kennedy2025analysis}.
Figure~\ref{figResonance} also shows how sensitive the resonance is to the variation of $E_\JJ^\atom$.

In any case, we expect that one can recover the resonance by tuning the atom transition frequency.
It can be varied in a rather wide range \emph{in situ}, by changing the atom bias $\Phi_\ext$.
That is shown in the Fig.~\ref{figResonance} lower panel.

Note that a change in $E_\JJ^\atom$ also shifts the resonator frequency $\omega_\res$ as shown in the Fig.~\ref{figResonance} lower panel.
That occurs as the atom equilibrium $\phi_\atom^\mi$ shifts (see the equations in Appendix~\ref{apEquilibrium}), which changes the coupler static phase difference $\delta$.
In its turn, this alters the magnitude of the coupler Josephson inductance that loads the resonator [see Eq.~\eqref{eqRenormELr}].
A similar contribution is present in the $\omega_\atom$ shift.
It arises due to the direct change of $E_\JJ^\atom$ in Eqs.~\eqref{eqRenormELa} and \eqref{eqPlasmas} combined with the associated change in $\phi_\atom^\mi$.

\subsection{Interactions}
\label{secRateEstimates}

\begin{figure}[t!]
\centering
\includegraphics{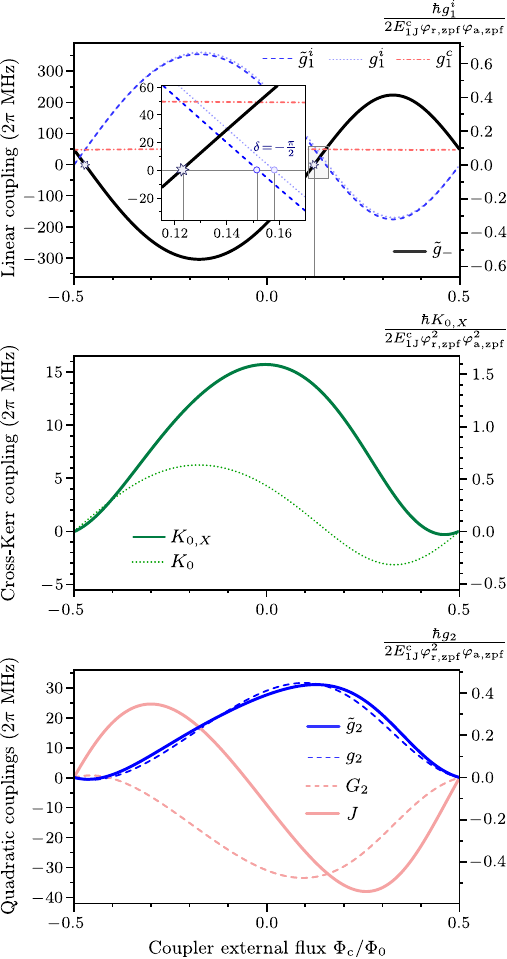}
\caption{Dependence of the interaction rates on the coupler bias $\Phi_\cpl$, as expressed in Table~\ref{tablBareInteractions}, Sec.~\ref{secInteractions}, and Appendix~\ref{apOtherInteractions}.
(Linear coupling): Total inductive $\tilde g^i_1$, bare inductive and capacitive $g^i_1$ and $g^c_1$, total excitation-number--preserving $\tilde g_- = g_1^c - \tilde g_1^i$.
The inset shows the vicinity of $\Phic$ where either $g^i_1$ or $\tilde g^i_1$ become zero (circles).
Stars mark the biases where $g_-$ vanishes.
(Cross-Kerr coupling): $K_{0,X}$ [Eq.~\eqref{eqCrossKerrMainPart}] with the asymmetry-induced correction, bare $K_0$.
(Quadratic couplings): Full reverse optomechanical $J$, two-photon bare $g_2$, full two-photon $\tilde g_2$.
The coupling quadratic in the artificial atom phase has the rate $G_2$.
rf SQUID regime as in Fig.~\ref{figTwoWells}.
Dimensionless axes neglect dependence of the zero-point amplitudes of the resonator and atom, $\varphi_{\res,\zpf}$ and $\varphi_{\atom,\zpf}$, on the coupler Josephson energy as it varies with $\Phi_\cpl$.
The maximal magnitude of the coupler Josephson energy is denoted by $2E_{1\JJ}$.
}
\label{figCouplings}
\end{figure}

Figure~\ref{figCouplings} demonstrates how the interactions rates depend on the coupler external flux $\Phi_\cpl$.
In addition to the trivial modulation of the coupler Josephson energy, $\Phi_\cpl$ also alters the symmetry of the Josephson interaction.
That is, it regulates the onset of the odd components in the coupling energy, including the two-photon interaction.
For each value of $\Phi_\cpl$ in the plots, we numerically solve the equations for equilibrium point in Appendix~\ref{apEquilibrium}, to obtain $\phi^\mi_\res$ and $\phi^\mi_\atom$.

\paragraph*{The top panel} in Fig.~\ref{figCouplings} shows the full single-photon coupling rate $\tilde g_- = g_1^c - \tilde g_1^i$ along with the two contributions that constitute it: the capacitive coupling $g^c_1$ as in Table~\ref{tablBareInteractions} and the full inductive coupling $\tilde g^i_1$ as in Eq.~\eqref{eqInductiveCouplingRenorm}.
We also plot the \emph{bare} inductive coupling $g^i_1$ as in Table~\ref{tablBareInteractions}.
Notice that the capacitive coupling is virtually constant in the coupler bias.
That is to be expected, as renormalizations due to the coupler Josephson inductance in Eqs.~\eqref{eqRenormELr} and \eqref{eqRenormELa} are negligible for $E^\cpl_\JJ \ll E^\res_L, E^\atom_L$.

By setting $\Phic\!\mod\Phi_0 = \Phi_0/2$, we turn off the coupler Josephson energy, $E^\cpl_\JJ = 0$.
Thus, all linear and nonlinear inductive-type couplings vanish, leaving only the capacitive coupling.
On the other hand, the bare inductive coupling $g^i_1$ vanishes as $\Phic$ satisfies $\delta\!\mod \pi = \pi/2$, where $\delta$ is given by Eq.~\eqref{eqCouplerPhase}.

The full inductive coupling $\tilde g^i_1$ has an additional component due to the asymmetry in the atom potential.
With the cubic nonlinearity coefficient $\mu = 1.651$ (for $E_\JJ^\cpl = 0$ and $C'_\cpl = 0$), this component is noticeable in the plot as a difference between $\tilde g_1^i$ and $g_1^i$.
As follows from Eq.~\eqref{eqInductiveCouplingRenorm}, $\tilde g^i_1$ vanishes for the bias determined by the condition $\cot \delta = \tfrac{5}{6} \mu\varphi_{\atom,\zpf}^2$.
The zero-shifting component $\tilde g_i - g_i$ is proportional to the bare $G_2$ interaction (see lower panel) and $\sin \delta$.
Hence, this component is close to its maximal magnitude right near the zero of the bare coupling $g^i_1$.
Expression of the full rate $\tilde g_-$ suggests an important property: for specific values of the external flux $\Phic$, the capacitive and inductive couplings cancel each other out, and the single-photon coupling switches off~\cite{neumeier2013single,ciani2019hamiltoncompcpbcrosskerr, kounalakis2018tuneable,sank2025balanced}.
We mark these values of $\Phic$ in Fig.~\ref{figCouplings}.

\paragraph*{The middle panel} in Fig.~\ref{figCouplings} demonstrates the cross-Kerr coupling rate $K_{0,X}$ as in Eq.~\eqref{eqCrossKerrMainPart} that takes into account virtual processes due to the atom cubic nonlinearity.
We also plot the bare rate $K_0$ as in Table~\ref{tablBareInteractions};
as follows from the equation, it vanishes for $\Phic$ at either $\delta \mod \pi = \pi/2$ or $\Phic\!\mod\Phi_0 = \Phi_0/2$ analogously to the inductive coupling $g^i_1$.

Note that we do not tune the resonance for every $\Phic$.
According to Fig.~\ref{figResonance}, the two-photon coupling becomes off-resonant further away from $\Phic = 0$.
Hence an additional cross-Kerr coupling arises.
However, it is as small as the other dispersive couplings neglected in Sec.~\ref{secCrossKerr}.

\paragraph*{The bottom panel} in Fig.~\ref{figCouplings} shows the couplings with the energy quadratic in the atom variables, particularly the sought-after two-photon coupling.
The bare two-photon coupling $g_2$ dominates the corrections in $\tilde g_2$ as provided in Appendix~\ref{apTwoPhotonFormula} and illustrated in Fig.~\ref{figDiagramsTwoPhoton}.
As in the cross-Kerr coupling rate, the corrections are more pronounced with higher linear coupling strength $g_\pm$.
Close to the maximal magnitude of the two-photon coupling, $\tilde g_\pm$ is vanishing and the two-photon corrections in Appendix~\ref{apTwoPhotonFormula} are not even visible.

This implies that the maximal strength of the two-photon coupling $\tilde g_2$ depends roughly linearly on the coupling energy $E^\cpl_\JJ$ according to Table~\ref{tablBareInteractions}.
On the other hand, the value of $E^\cpl_\JJ$ is bounded by the condition~\eqref{eqSizableEJcNoUltrastrong}.
When it does not hold, our perturbative treatment of the off-resonant single-photon coupling breaks down.

Note, that for the unbiased coupling SQUID, when $\Phic = 0$, the magnitude of the two-photon coupling is only about 11\% smaller than at its maximum.
Hence, if that is a reasonable tradeoff, a single Josephson junction may be used instead of a coupling SQUID, possibly reducing the amount of noise in the system.
In that case, two-photon coupling arises solely due to a nonzero equilibrium phase of the metastable rf SQUID, as discussed in the preceding work~\cite{parti}.

\section{Discussion}
\label{secDiscussion}

Here we discuss the bosonic approximation, neglecting some nonlinearities, coupling to the environment, and whether our correction are quantum.

\subsection{Commutation relations for the atom operators}

\paragraph{Bosonic approximation.}
In determining the corrections, we rely on the bosonic approximation for the artificial atom metastable states.
In other words, we assume that the atom does not leave its metastable well as in Fig.~\ref{figTwoWells};
while, at the same time, it can be excited arbitrarily high in energy.
Clearly, these two assumptions are contradictory for a metastable potential.
However, we expect them to hold for the atom phase dwelling sufficiently close to the metastable well minimum. 

More precisely, we assess the plausibility of the bosonic approximation by looking into the virtual processes that comprise the corrections.
We have found that the main cross-Kerr renormalization in Eq.~\eqref{eqCrossKerrMainPart} requires at least three coherent energy states as shown for a contributing virtual process in Fig.~\ref{figDiagramsLinearAndCrossKerr}(I).
Other less important corrections require at most four coherent energy states to be available:
See Fig.~\ref{figDiagramsLinearAndCrossKerr}(II) for a single-photon $(n{+1})g \leftrightarrow ne$ correction as in Eqs.~\eqref{eqInductiveCouplingRenormHalfBaked}--\eqref{eqInductiveCouplingRenorm} and Fig.~\ref{figDiagramsTwoPhoton}(b) for a two-photon $(n{+2})g \leftrightarrow ne$ correction as in Appendix~\ref{apTwoPhotonFormula}.

All mentioned processes occur well within the seven metastable energy states of the rf SQUID in Sec.~\ref{secEstimates}.
On the other hand, a process going beyond that, occurs differently than predicted by the bosonic approximation.
Such real process can bring the system out of the metastable state subspace as the rf SQUID relaxes into the right well in Fig.~\ref{figTwoWells}.
At the same time, a virtual process as predicted by the bosonic model may not occur in that case---if the dissipation into the right well is too fast.

\paragraph{Quasispin approximation as an alternative.}
Another approach is to cut the Hilbert space to the metastable states right from the start~\cite{sokolov2020superconducting,sokolov2023thesis}.
In this model, any process exciting the atom further above simply does not occur;
algebraically, for example, that corresponds to $\sigma_+ \ket u = 0$ for a system with the highest energy state $\ket u$, where the raising operator acts as $\sigma_+ \ket g = \ket e$ etc.
This model predicts that a real process going above the metastable levels does not occur, although in reality the system can leave the metastable subspace in that case.
However, for the calculation of renormalizations due to virtual processes, quasispin approximation may be a better model as it takes into account only the processes within the quasispin subspace.
This restriction from above is encoded into the commutation relations $[\sigma_+, \sigma_-] = Z$.
Here $\sigma_+ = (X+Y)/2$ and $\sigma_- = (X-Y)/2$ are the raising and lowering operators, where $\hbar X/2$, $\hbar Y/2$, and $\hbar Z/2$ are the quasispin projection operators for the respective axes.
In other words, in $\sigma_- \sigma_+ = \sigma_+\sigma_- - Z$, the state-dependent addition $Z$ on the right-hand side ensures that it vanishes when acting on the highest excited state $\ket u$.

We only mention here that our calculations in Appendix~\ref{apRenormHamiltonianDerivation} would have become even more tedious in this approach, where the commutator of the ladder operators is operator-valued.
As for the question of determining the best model to describe metastable energy states, we leave it for future work.
A similar question arises for modeling transmons which can ``ionize'' due to energetic driving~\cite{shillito2022dynamics,dumas2024ionization}.
Even putting ionization aside, transmons are typically modeled analytically as anharmonic oscillators~\cite{marxer2023long,koch2007transmon,blais2021circuit}, and numerically as a quasispin~\cite{marxer2025fidelity}.

\paragraph{Two-photon coupling.}
As argued in Sec.~\ref{secTwoPhotonIsBare}, the main part of the two-photon coupling rate $\tilde g_2$ typically comes from the bare interaction $g_2$.
That is the case for reasonable linear coupling rates and comparable zero-point fluctuations of the atom and resonator.
We have also demonstrated that in the estimates of Sec.~\ref{secEstimates}.
That confirms our assumption in the preceding work~\cite{parti}, where we claimed it is enough to assess the bare two-photon coupling.
Moreover, that suggests that our estimate for attainable magnitudes of the two-photon coupling does not rely on the bosonic approximation.

\subsection{Neglected resonator anharmonicity}
\label{secDiscussionAnharmonicity}

In tuning the resonance in Secs.~\ref{secResonatorParameters} and \ref{secResonance}, we neglect the shift $\varXi_\res$ in the resonator frequency, which is of the same magnitude as the induced resonator anharmonicity with the energy $(-\varXi_\res/2) a^{\dag2} a^2$.
This resonator anharmonicity (self-Kerr nonlinearity) arises due to the coupler nonlinear Josephson inductance parallel to the resonator inductance.
Note that the resonator anharmonicity also dresses owing to the nonlinear interactions.
As this part does not enter the population-independent frequency shift, we are not interested in it in this work.

First, let us assess the anharmonicity magnitude for $\Phic = 0$, i.e., close to the maximal values of the two-photon coupling.
According to Appendix~\ref{apSelfNonlinearitiesTreatment}, the bare part of the resonator anharmonicity is $\varXi_\res \approx \varXi_{\res3} + \varXi_{\res4}$ with $\varXi_{\res3} = 60X_\res^2 / \omega_\res^p$ and $\varXi_{\res4} = 12Y_\res$.
The latter denote the contributions from the cubic and the quartic parts of the resonator potential.
The potential is written down in Appendix~\ref{apCircuitHamiltonian}.
Using the expressions therein, we calculate $X_\res / (\cos\frac{\pi\Phic}{\Phi_0} \sin\delta) \approx 2\pi\,11$\,MHz and $Y_\res / (\cos\frac{\pi\Phic}{\Phi_0} \cos\delta) \approx 2\pi\,0.36$\,MHz with the parameters from Sec.~\ref{secEstimates}.
Then, approximating $\delta \approx -\phi_\atom^\mi - \pi\Phi_\cpl / \Phi_0$, we obtain $\varXi_{\res3} \approx 2\pi\,1.0$\,MHz and $\varXi_{\res4} \approx 2\pi\,1.2$\,MHz for $\Phic = 0$.

In fact, we find that the condition $\varXi_\res < g_2$ qualifies when $(g_2)_{\delta = \pi/2} < \frac{3\omega_\atom}{20\sin\delta} \big(\frac{\varphi_{\atom,\zpf}^2}{\varphi_{\res,\zpf}^2} + \varphi_{\atom,\zpf}\cot\delta \big)$.
Hence we expect the two-photon resonance to be unaffected by the induced resonator anharmonicity under normal conditions.
Namely, that is the case below ultrastrong coupling threshold as in Eq.~\eqref{eqSizableEJcNoUltrastrong} as long as $\varphi_{\atom,\zpf}^2 / \varphi_{\res,\zpf}^2 \gtrsim 1$.

\subsection{Other effects of the coupling SQUID}

At first glance, the Josephson energy $E_\JJ^\cpl$ of the coupler only gives rise to inductive interactions with the strengths proportional to $E_\JJ^\cpl$.
Unlike the preceding part of this work~\cite{parti}, we take into account how $E_\JJ^\cpl$ influences other system parameters.

First of all, the coupler Josephson inductance loads inductances of the resonator and atom.
That shifts the transition frequencies as in Secs.~\ref{secCouplerParameters} and \ref{secResonatorParameters}.
It also alters the anharmonicities as in Sec.~\ref{secResonatorParameters} and \ref{secDiscussionAnharmonicity}.

The coupling rates also depend on $E_\JJ^\cpl$ via the magnitudes of the zero-point fluctuations.
For $10E_\JJ^\cpl \sim E_L^\res, E_L^\atom, E_\JJ^\atom$, the variation in a coupling of order $n$ in the zero-point fluctuations can be up to $(n/4) \times 10\%$, according to Eqs.~\eqref{eqResZPF}, \eqref{eqAtomZPF}, and \eqref{eqInductiveCouplingRenorm}.

That also shifts the maximum of the two-photon coupling in Fig.~\ref{figCouplings}.
In addition, the maximum is influenced by the related shift of the equilibrium point.
As follows from the equilibrium equations in Appendix~\ref{apEquilibrium}, an additional current through the coupler shifts its static phase $\delta$ depending on the bias $\Phic$.
We calculate (results not presented), that this variation in $\delta$ is about $0.1\pi$ for the parameters as in Sec.~\ref{secEstimates}.
That may be further analyzed in future work, along the lines of Refs.~\cite{geller2015tunable,li2011decoherence}.

\subsection{Expansion of the Josephson coupling energy}
\label{secDiscussionNonlinearities}

It is possible to find the coefficients for the Hamiltonian terms as in Appendix~\ref{apSecondQuantizedHamiltonian} and higher in all orders of the coupling energy expansion~\cite{trif2015photon,ma2022photon,baskov2025exact}.
However, we resort to the fourth-order expansion in our starting Hamiltonian.
The reason is that we work with a relatively strong Josephson coupling to boost the two-photon processes.
Consider the perturbative two-photon corrections in Appendix~\ref{apTwoPhotonFormula}.
They are of the same order in the small zero-point fluctuations as the neglected higher-order expansion terms in the coupler energy.
However, with stronger Josephson coupling, these corrections exceed the higher-order expansion terms and should be accounted for.
With the relatively strong Josephson coupling, we need to tackle even higher perturbative corrections as we include the higher expansion terms.
That is beyond the scope of this work.

\subsection{Coupling to environment}

If both the resonator and the artificial atom couple to the same or correlated heat baths, an additional linear coupling arises~\cite{metelmann2015nonreciprocal, wang2020dissipative}.
That influences the small two-photon rate dressing.
Indeed, the latter depends on the linear coupling rate, as shown in Fig.~\ref{figDiagramsTwoPhoton} and Appendix~\ref{apTwoPhotonFormula}.
However, such contributions are small in the strong coupling regime, where the interactions are faster than the dissipation.

\subsection{Quantumness}

Mathematically, the vacuum loops in Figs.~\ref{figDiagramsLinearAndCrossKerr} and \ref{figDiagramsTwoPhoton} arise from the commutation relations between the operators.
Therefore, it may seem, that every correction there is a quantum effect---since each of them contains a loop.

However, that is not the case for the corrections with a single vacuum loop.
We demonstrate this for the main cross-Kerr correction, by obtaining it from the dynamical equations in Appendix~\ref{apDynamicalEqs}.
This derivation is valid in the classical limit with the complex phasors~\cite{zakharov1985hamiltonian,sokolov2024signatures} instead of the operators $a$ and $b$.
Hence, our single-loop corrections are also present in the classical systems.

Roughly speaking, one loop in the diagrams and the corresponding commutator arise as the Schrieffer-Wolff transformation takes into account a part of the system dynamics.
Similarly to the commutators we use to obtain the Heisenberg equations of motion from a Hamiltonian, the corresponding commutators do not signify a quantum effect.

\section{Uses of the proposed circuit and its generalizations}
\label{secApplications}

\subsection{Two-photon detection}

A phase qubit with a two-well potential, similar to Fig.~\ref{figTwoWells}, can act as a Josephson photomultiplier~\cite{opremcak2018measurement,opremcak2020high,shnyrkov2023rfsquid,ilinskaya2024fluxqubit}.
After such device absorbs the resonator photon(s), it switches to the true ground state, which indicates the presence of photon(s).
One may use two-photon absorption to detect photon pairs.

Consider a device where the excited metastable state tunnels, immediately on absorption, to the deeper well~\cite{chen2011microwave,govia2012theory,schoendorf2018optimizing,sokolov2020superconducting}.
To achieve fast and incoherent tunneling, the excited state may lie close to the barrier top while the states above and on the other side of the barrier are extremely dissipative.
With only two relatively coherent metastable states, our bosonic approximation may break as discussed in Sec.~\ref{secDiscussion}.
The theory as in Refs.~\cite{sokolov2020superconducting,sokolov2023thesis} may be more appropriate in that case.

Still, we expect that the two-photon coupling rate is $\tilde g_2 \approx g_2 \approx 2\pi \, 30$\,MHz approximately as in Fig.~\ref{figCouplings}.
This approximate equality is valid even if the small dressing as predicted by the bosonic theory vanishes.
As the rate $g_2$ is below the typical tunneling rates of 50--100\,MHz, it limits the detection of a photon pair.
Hence the detection time is about $1 / g_2 \sim 10$\,ns, comparable to the related device proposed in Ref.~\cite{stolyarov2025detector} with one of the authors.

\subsection{Quantum-nondemolition qubit readout}
\label{secQNDReadout}

We can suppress the single-photon coupling $\tilde g_-$ while retaining the cross-Kerr one.
With purely inductive $g_- = g_1^i$, the \emph{bare} cross-Kerr $K_0$ vanishes for the same bias as the bare linear coupling $g_-$, i.e., at $\delta \mod \pi = \pi/2$.
To retain the cross-Kerr interaction while switching off the linear coupling, in Ref.~\cite{neumeier2013single} the capacitive coupling shifts the zero of the linear coupling rate as we illustrate in Fig.~\ref{figCouplings}.
In our case, near the zero of $\tilde g_-$, the cross-Kerr interaction arises due to asymmetry of the atom potential, even without the capacitive coupling.

As a result, it can be used for qubit readout, similarly to Refs.~\cite{dassoneville2020fast,chapple2025balancedcrosskerr,beaulieu2026fast,wang2025longitudinal}.
In Fig.~\ref{figCouplings}, the cross-Kerr coupling rate $K_{0,X}$ is about 14\,MHz where the linear coupling $\tilde g_-$ vanishes.
We also need to avoid two-photon exchange.
Consider a detuning of $\omega_\atom - 2\omega_\res = 2\pi\,2$\,GHz from the two-photon resonance for the two-photon coupling $g_2 = 2\pi\,30$\,MHz.
In that case, the readout is approximately quantum-nondemolition~\cite{braginsky1996quantum,haroche1999cavity,blais2021circuit,sillanpaa2005direct}, as the cross-Kerr coupling term in the Hamiltonian~\eqref{eqTwoPhotonHamiltonian} commutes with the qubit part.
In our estimate, we neglect the contribution of $g_2^2 / (\omega_\atom - 2\omega_\res) \sim 2\pi\,1$\,MHz to the cross-Kerr rate stemming from the nonresonant two-photon interaction.

Consider using a protocol in Ref.~\cite{chapple2025balancedcrosskerr} with the resulting cross-Kerr coupling of $K_{0,X} \approx 2\pi\,16$\,MHz.
To estimate the readout fidelity $\mathcal{F}$, we assume that the qubit lifetime is much longer than the measurement time.
Then, according to Ref.~\cite{chapple2025balancedcrosskerr}, $\mathcal{F} = 1 - \frac12 \mathrm{erfc}(\mathrm{SNR}/\sqrt{8})$, where $\mathrm{erfc}$ is the complementary error function.
The signal-to-noise ratio (SNR) of the integrated output signal is~\cite{mori2025highpowerreadout, blais2021circuit, gambetta2008quantum}
\begin{equation} \label{eqSNR}
   \mathrm{SNR} = \sqrt{4 \eta \bar n_\mathrm{m} \tau_\mathrm{m} \Gamma_\res \frac{K_{0,X}^2}{\Gamma_\res^2 + K_{0,X}^2}}
\end{equation}
for a measurement of time $\tau_\mathrm{m}$.
Let us assume there is, on average, $\bar n_\mathrm{m} = 30$ photons in the probe impulse.
To maximize the SNR, we choose the resonator linewidth $\Gamma_\res = K_{0,X}$.
Consider the quantum efficiency of the readout chain $\eta = 30\%$, which is accessible using near-quantum-limited amplifiers.
With the cross-Kerr coupling $K_{0,X}$ as above, the phase qubit in Sec.~\ref{secEstimates} can be read out in $\tau_\mathrm{m} \approx 30$\,ns with over $99.9\%$ fidelity.
For a low efficiency $\eta=5\%$, on par with a setup with no quantum limited-amplifier~\cite{beaulieu2026fast}, fidelity reaches $99.9\%$ in 150\,ns.

Besides the phase qubit, such readout should work for other qubits with an asymmetrical potential.
Our theory can also be useful for assessing other~\cite{ciani2019hamiltoncompcpbcrosskerr,neumeier2013single} applications of the cross-Kerr coupling.

\section{Conclusions}
\label{secConclusions}

We have calculated dressing of the resonant two-photon and cross-Kerr couplings mediated by a symmetrical SQUID.
In a systematic approach, we expand the nonlinear Josephson coupling to the fourth order and account for the leading second-order perturbative corrections.
With diagrams, we illustrate some correction processes and elucidate the minimum number of the metastable energy states required for our quantitative results to hold.

We argue that the two-photon coupling normally acquires only small corrections, even for higher energies of the Josephson coupling.
On the other hand, the cross-Kerr coupling can obtain the most significant renormalization.
It arises due to nonlinear coupling and asymmetry in one of the coupled system potentials.
For a single-mode resonator coupled to a phase qubit, this correction can reach 15\,MHz as estimated in Sec.~\ref{secEstimates}.
While the bare cross-Kerr coupling can be avoided with a proper coupler bias, this dressing part only vanishes with the bias switching off all Josephson-mediated interactions.

We have discussed the validity of our theory in Sec.~\ref{secDiscussion}.
Most importantly, we have used the bosonic approximation, i.e., we model the artificial atom as an anharmonic oscillator near its metastable minimum.
Our quantitative results on the renormalizations depend on that model validity.
In Sec.~\ref{secEstimates}, we have provided estimates for an example system.
They can be verified with further numerical or experimental studies.

However, we have verified that the virtual processes in the important corrections excite the atom at most to the third excited metastable state.
Hence we expect that our theory should hold reasonably well for a phase qubit with seven coherent metastable states.
To count the required number of energy states, we analyze the Schrieffer-Wolff transformations used.
This approach may also be used for transmons, whose excitations are bosonic only within a limited number of energy states.

With modifications, the considered circuit can be applied for two-photon detection and qubit readout.
If relatively high rates of the cross-Kerr coupling are tolerable or required, our nonlinear coupler may be more favorable than more complex ones, such as SNAIL~\cite{frattini2017threewave,hillmann2022designingkerr}, by the fact that it requires fewer junctions.
That is especially true when it can be simplified further into a single junction as discussed in Sec.~\ref{secRateEstimates}.

\begin{acknowledgments}

A.~M.~S.\ and E.~V.~S.\ thank F.~Wilhelm-Mauch for discussing our early concepts and results, and R.~Baskov for discussions and useful comments.
E.~V.~S.\ thanks A.~Semenov for discussions and useful comments.
E.~V.~S.\ acknowledges support from the National Research Foundation of Ukraine through the project No.\ 2023.03/0165, Quantum correlations of electromagnetic radiation.
A.~M.~S.\ acknowledges partial support from the Academy of Finland (Contract No.\ 321982) and the President of Ukraine scholarship for young scientists.

\end{acknowledgments}

\appendix

\section{Circuit Hamiltonian}
\label{apCircuitHamiltonian}

In this appendix, we provide our starting model Hamiltonians for the circuit in Fig.~\ref{figCircuit}.
We also write down equations to determine the system equilibrium.

\subsection{Fourth-order expansion of Josephson energy}

We derived the general circuit Hamiltonian in the preceding work~\cite{parti}, where the resulting expressions are given in Eqs.~(1) and (4)--(6).
Now we expand it around equilibrium, up to the fourth order in $\varphi_\res$ and $\varphi_\atom$.
Then, we arrive at the quantized Hamiltonian
\begin{equation}
\label{eqQuantizedHamiltonian}
	\opH = \opH_\res + \opH_\atom + \opH_\cpl,
\end{equation}
by imposing the usual commutation relation $[\varphi_\res, n_\res] = i$ for the resonator variables and the analogous one for the rf SQUID.
In the Hamiltonian in Eq.~\eqref{eqQuantizedHamiltonian}, the resonator, atom, and the coupling parts read
\begin{widetext}
\begin{align}
\label{eqHrFirstQuantized}
	\opH_\res \approx {}&
		4 E^\res_C n^2_\res + \frac 1 2 \tilde{E}^\res_L \varphi_\res^2
			- \frac 1 6 E^\cpl_\JJ \varphi_\res^3 \sin \delta
			- \frac 1{24} E^\cpl_\JJ \varphi_\res^4 \cos \delta,
\\
\label{eqHaFirstQuantized}
	\opH_\atom \approx {}&
		4E^\atom_C n^2_\atom + \frac 1 2 \tilde{E}^\atom_L \varphi_\atom^2
			- \frac 1 6 (E^\atom_\JJ \sin \phi^\mi_\atom
			- E^\cpl_\JJ \sin \delta)\varphi^3_\atom
			- \frac 1{24} (E^\atom_\JJ \cos \phi^\mi_\atom
			+ E^\cpl_\JJ \cos \delta)\varphi_\atom^4,
\\
\label{eqHcFirstQuantized}
	\opH_\cpl \approx {}& 
		E^\cpl_C n_\atom n_\res
		- E^\cpl_\JJ \varphi_\res \varphi_\atom \cos \delta
\nonumber
\\	
		& {}+ \frac 1 2 E^\cpl_\JJ (\varphi_\res^2 \varphi_\atom
		- \varphi_\res \varphi_\atom^2) \sin \delta
		- \frac 1 4 E^\cpl_\JJ \varphi_\res^2 \varphi_\atom^2 \cos \delta
		+ \frac 1 6 E^\cpl_\JJ (\varphi_\res^3 \varphi_\atom
		+ \varphi_\res \varphi_\atom^3) \cos \delta.
\end{align}
\end{widetext}
Each part consists of the ``kinetic'' energy contribution quadratic in $n_\res$ and $n_\atom$, the potential energy terms quadratic in $\varphi_\res$ and $\varphi_\atom$, and the higher-order terms that give rise to self-nonlinearities and nonlinear interactions.

Compared to the preceding work~\cite{parti}, we rearrange the terms in the Hamiltonians in Eqs.~\eqref{eqHrFirstQuantized}--\eqref{eqHcFirstQuantized}.
Here the resonator and the atom Hamiltonians accommodate the self-nonlinearities induced by the coupling SQUID.

\subsection{Equilibrium point}
\label{apEquilibrium}

Formally, one can find the equations for the equilibrium point from the full dynamical equations.
The full Hamiltonian, as provided in the preceding work~\cite{parti}, generates such full equations.
Then, setting the derivatives of the phase and charge variables to zero, one obtains
\begin{subequations} \label{eq:mins}
\begin{gather}
\label{eqResonatorExtremum}
	E^\res_L \phi^\mi_\res + E^\cpl_\JJ \sin \delta = 0,
\\
\label{eqAtomExtremum}
	E^\atom_L \Big(\phi^\mi_\atom
					 - \frac{2\pi\Phie}{\Phi_0}\Big)
	+ E^\atom_\JJ \sin\phi^\mi_\atom - E^\cpl_\JJ \sin \delta
	= 0,
\end{gather}
\end{subequations}
where $\Phie$ is the rf SQUID bias flux and the static phase difference $\delta$ is defined in Eq.~\eqref{eqCouplerPhase}.
As discussed in Ref.~\cite{parti}, the equations express the conservation of static currents.
Alternatively, Eq.~\eqref{eqResonatorExtremum} can be obtained by finding the screening current induced by the external magnetic field.

\subsection{Hamiltonian with ladder operators}
\label{apSecondQuantizedHamiltonian}

We rewrite the Hamiltonian in Eqs.~\eqref{eqQuantizedHamiltonian}--\eqref{eqHcFirstQuantized} in terms of the ladder operators introduced in the main text.
Hamiltonians of the resonator and artificial atom read
\begin{gather}
\label{eqHrBare}
	\frac{\opH_\res}{\hbar} = \omega^p_\res \Big(a^\dag a + \frac 1 2\Big)
				- X_\res (a^\dag + a)^3 - Y_\res (a^\dag + a)^4,
\\
\label{eqHaBare}
	\frac{\opH_\atom}{\hbar} = \omega^p_\atom \Big(b^\dag b + \frac 1 2\Big)
				- X_\atom (b^\dag + b)^3 - Y_\atom (b^\dag + b)^4.
\end{gather}
The coupling Hamiltonian in Eq.~\eqref{eqQuantizedHamiltonian} becomes
\begin{align}
\label{eqHcBare}
	\frac{\opH_\cpl}{\hbar} =
		&-g^c_1 (a^\dag - a) (b^\dag - b) - g^i_1 (a^\dag + a)(b^\dag + b)
\nonumber
\\
		&{}-\frac{K_0}{4} (a^\dag + a)^2(b^\dag + b)^2
\nonumber
\\
		&{}- \sum_{n=2,3} g_n (a^\dag + a)^n (b^\dag + b)
\nonumber
\\
		&{}- \sum_{n=2,3} G_n (a^\dag + a)(b^\dag + b)^n.
\end{align}
Apart from the bare interactions in Table~\ref{tablBareInteractions}, the Hamiltonian describes the three-photon and three--atom-excitation interactions with rates
\begin{gather}
\label{eqBareThreePhotonCoupling}
  	g_3 = -\frac{E^\cpl_\JJ}{6\hbar}
		\varphi_{\res,\zpf}^3 \varphi_{\atom,\zpf} \cos\delta,
\\
\label{eqG3}
	G_3 = -\frac{E^\cpl_\JJ}{6\hbar}
		\varphi_{\res,\zpf} \varphi_{\atom,\zpf}^3 \cos\delta.
\end{gather}
They take part in lesser renormalizations as derived in Appendix~\ref{apTwoPhotonDerivation} and discussed in Appendix~\ref{apOtherInteractions}.

In addition to the cubic nonlinearity in the atom potential as in Table~\ref{tablBareInteractions} in the main text, we provide other self-nonlinearities in the Hamiltonians~\eqref{eqHrBare} and \eqref{eqHaBare}.
Quartic term in the atom potential with strength
\begin{equation}
\label{eqYaFull}
	Y_\atom =
		\frac{E^\atom_\JJ \cos\phi_\atom^\mi + E^\cpl_\JJ\cos\delta}
			 {24\hbar}
		\varphi_{\atom, \zpf}^4
\end{equation}
can influence its transition frequency and anharmonicity as estimated in Sec.~\ref{secEstimates}.
Similarly to $X_\atom$ in Eq.~\eqref{eqXaFull}, one can rewrite it as
\begin{equation}
\label{eqYaZPF}
	Y_\atom = \frac{\eta}{48} \omega_\atom^p \varphi_{\atom,\zpf}^2
\end{equation}
with $\eta = (\tilde E^\atom_L - E^\atom_L)/\tilde{E}^\atom_L \approx \cos\phi_\atom^\mi / (\cos\phi_\atom^\mi + \beta_L^{-1})$.
The resonator self-nonlinearities
\begin{equation}
\label{eqResonarorSelfNonlinearities}
\begin{split}
	X_\res = \frac{E^\cpl_\JJ}{6\hbar} \varphi_{\res,\zpf}^3 \sin\delta,
\quad
	Y_\res = \frac{E^\cpl_\JJ}{24\hbar} \varphi_{\res,\zpf}^4 \cos\delta
\end{split}
\end{equation}
arise solely due to the coupling SQUID nonlinearity.
They give rise to the resonator anharmonicity as estimated in Sec.~\ref{secDiscussionAnharmonicity}.

\section{Derivation of the renormalized two-photon Hamiltonian}
\label{apRenormHamiltonianDerivation}

In this appendix, we reduce the approximate circuit Hamiltonian in Appendix~\ref{apCircuitHamiltonian} to a variation of the two-photon Jaynes-Cummings model where the artificial atom is modeled as an anharmonic oscillator.
First, we show how the nonlinear character of the atom potential gives rise to the cross-Kerr interaction.
Then, we calculate renormalizations in the anharmonicities and two-photon coupling, originating from the nonresonant interactions.

\subsection{Eliminating the cubic perturbation}
\label{apSelfNonlinearitiesTreatment}

We start with treating the cubic self-nonlinearities in the Hamiltonian from Appendix~\ref{apSecondQuantizedHamiltonian} as a perturbation.
We eliminate them from the Hamiltonian~\eqref{eqHaBare} with a Schrieffer-Wolff transformation \cite{zhu2013cqedfluxonium,blais2004cavity,bravyi2011,lamberto2025renormalization}, $\opH_\atom \rightarrow e^{-\Lambda_\atom \opS_\atom} \opH_\atom e^{\Lambda_\atom \opS_\atom}$, where
\begin{equation} \label{eqLambdaAtomDefinition}
	\Lambda_\atom = \frac{X_\atom}{\omega^p_\atom}.
\end{equation}
The anti-Hermitian operator $\opS_\atom$ is given by \cite{hillmann2022designingkerr, stolyarov2023pnr}
\begin{equation} \label{eqAtomTransformation}
	\opS_\atom = \frac{1}{3}(b^{\dag 3} - b^3) + 3(b^{\dag 2} b - b^\dag b^2) + 3(b^\dag - b).
\end{equation}
For manual calculations, one may rewrite $\opS_\atom$ in terms of $b^\dag \pm b$ as given in Appendix~\ref{apWick}.
Here, we use the form as in Eq.~\eqref{eqAtomTransformation} and perform the tedious calculations with the computer algebra package \textsc{QuantumAlgebra.jl}~\cite{qalg}.

Using the Baker-Campbell-Hausdorff (BCH) formula up to the second commutator,
\begin{equation} \label{eqBCH}
	e^{-\Lambda_\atom \opS_\atom} \opH_\atom e^{\Lambda_\atom \opS_\atom}
		\approx \opH_\atom + \Lambda_\atom [\opH_\atom, \opS_\atom]
		+ \frac{\Lambda^2_\atom}2 \big[[\opH_\atom,\opS_\atom],\opS_\atom\big],
\end{equation}
we arrive at the Hamiltonian of a quantum anharmonic oscillator,
\begin{equation}
\label{eqAtomHamiltonian}
	\frac{\opH_\atom}{\hbar} \approx \omega^p_\atom b^\dag b - \varXi_\atom b^\dag b - \frac{\varXi_\atom}{2} b^{\dag 2} b^2 + \omega^\zpf_\atom + V'_{\atom3} + V_{\atom4},
\end{equation}
up to the first order in $\Lambda_\atom$.
Here
\begin{equation}
\label{eqBareAtomAnharmonicityWithLambda}
	\varXi_\atom = 60 \Lambda_\atom X_\atom + 12Y_\atom
\end{equation}
denotes anharmonicity in the artificial atom energy levels
and $\omega^\zpf_\atom = \omega^p_\atom/2 - 11X_\atom^2 / \omega^p_\atom - 3Y_\atom$ is its zero-point energy in the frequency units.
The first term in Eq.~\eqref{eqBareAtomAnharmonicityWithLambda} comes from the cubic perturbation in the atom Hamiltonian.
With Eq.~\eqref{eqBareAtomAnharmonicityWithLambda} for the anharmonicity, the Hamiltonian~\eqref{eqAtomHamiltonian} predicts the same corrections to the energy levels as the textbook calculation in Ref.~\cite{landau1991quantum}.

On several substitutions, we obtain the atom anharmonicity in terms of the circuit energies,
\begin{equation}
\label{eq:varXi_atom}
	\varXi_\atom = \frac{E^\atom_C}{\hbar}
	\Bigg[\frac53 \bigg(\frac{E^\atom_L}{\tilde{E}^\atom_L}\bigg)^2
	\Big(\phi^\mi_\atom - \frac{2\pi\Phie}{\Phi_0}\Big)^2
	+ \frac{\tilde E^\atom_L - E^\atom_L}{\tilde{E}^\atom_L} \Bigg].
\end{equation}
We have used Eqs.~\eqref{eqXaFull} and~\eqref{eqYaFull} for $X_\atom$ and $Y_\atom$, the atom plasma frequency in Eqs.~\eqref{eqPlasmas}, its full inductive energy in Eq.~\eqref{eqRenormELa}, zero-point fluctuations in Eq.~\eqref{eqAtomZPF}, and the equation on equilibrium~\eqref{eqAtomExtremum}.

For deriving Eq.~\eqref{eqAtomHamiltonian}, we have assumed that the cubic nonlinearity in the artificial atom is relatively weak with $|\Lambda_\atom| \ll 1$.
Apart from its definition Eq.~\eqref{eqLambdaAtomDefinition}, another way to express this small parameter is
\begin{equation}
\label{eqLambdaInTermsOfZPF}
	\Lambda_\atom = \frac{\mu}{12} \varphi_{\atom,\zpf},
\end{equation}
where $|\mu| \lesssim 2E^\atom_\JJ / \tilde E^\atom_L$ according to Eq.~\eqref{eq:def_mu} for $E_\JJ^\cpl \lesssim E_\JJ^\atom$.
With that, Eq.~\eqref{eqBareAtomAnharmonicityWithLambda} becomes
\begin{equation}
\label{eqBareAtomAnharmonicityWithZPF}
	\varXi_\atom = \frac{1}{4}\Big(\frac{5}{3}\mu^2 + \eta\Big)
		\omega_\atom^p \varphi_{\atom,\zpf}^2.
\end{equation}
where we have also used Eqs.~\eqref{eqXaFull} and \eqref{eqYaZPF}.

In the atom Hamiltonian~\eqref{eqAtomHamiltonian}, we collect the off-diagonal terms in the last two terms.
The penultimate term reads as
\begin{align}
	\label{eqOffDiagonal3}	
	V'_{\atom3} = & -18\Lambda_\atom X_\atom (b^{\dag 2} + b^2) - 12\Lambda_\atom X_\atom (b^{\dag 3} b + b^\dag b^3)
	\nonumber
	\\
	& + 3\Lambda_\atom X_\atom (b^{\dag 4} + b^4) - 136\Lambda_\atom Y_\atom(b^\dag + b)
	\nonumber
	\\
	& - 48\Lambda_\atom Y_\atom (b^{\dag 3} + b^3) - 336\Lambda_\atom Y_\atom (b^{\dag 2} b + b^\dag b^2)
	\nonumber
	\\
	&  + 8\Lambda_\atom Y_\atom (b^{\dag 5} + b^5) - 112\Lambda_\atom Y_\atom (b^{\dag 3} b^2 + b^{\dag 2} b^3)
	\nonumber
	\\
	& - 24 \Lambda_\atom Y_\atom (b^{\dag 4} b + b^\dag b^4)
\end{align}
and aggregates the terms arising upon transformation of the atom bare Hamiltonian~\eqref{eqHaBare}.
The last term in Eq.~\eqref{eqAtomHamiltonian} is the off-diagonal part of the atom quartic self-nonlinearity
\begin{equation}
	\label{eqOffDiagonalBare4}
	V_{\atom4} = -6Y_\atom (b^{\dag 2} + b^2) - 4Y_\atom (b^{\dag 3} b + b^\dag b^3) - Y_\atom (b^{\dag 4} + b^4).
\end{equation}

Just as for the atom, we apply a Schrieffer-Wolff transformation to the resonator Hamiltonian: $\opH_\res \rightarrow e^{-\Lambda_\res \opS_\res} \opH_\res e^{\Lambda_\res \opS_\res}$,
where $\Lambda_\res$ reads as
\begin{equation} \label{eqLambdaResDefinition}
	\Lambda_\res = \frac{X_\res}{\omega^p_\res}.
\end{equation}
Note that $|\Lambda_\res| \ll 1$: the resonator cubic nonlinearity is vanishingly small, as its only source is the SQUID coupler.
The operator $\opS_\res$ is obtained by replacing the atom operators with the resonator ones in Eq.~\eqref{eqAtomTransformation}.
With the same approximations as in Eq.~\eqref{eqAtomHamiltonian}, we arrive at
\begin{equation} \label{eqResonatorHamiltonian}
	\frac{\opH_\res}{\hbar} \approx \omega^p_\res a^\dag a - \varXi_\res a^\dag a - \frac{\varXi_\res}{2} a^{\dag 2} a^2 + \omega^\zpf_\res + V'_{\res3} + V_{\res4},
\end{equation}
where $\varXi_\res = 60\Lambda_\res X_\res + 12Y_\res$
is the resonator anharmonicity
and $\omega^\zpf_\res = \omega^p_\res/2 - 11X_\res^2 / \omega^p_\res - 3Y_\res$ is its zero-point energy in the frequency units.
We further transform the anharmonicity as
\begin{equation} \label{eq:varXi_res}
	\varXi_\res = \frac{E^\res_C}{\hbar}
		\Bigg[\frac 5 3 \Big(\frac{E^\res_L}{\tilde{E}^\res_L}
				\phi^\mi_\res\Big)^2
			  + \frac{\tilde E^\res_L - E^\res_L}{\tilde{E}^\res_L}\Bigg],
\end{equation}
analogously to Eq.~\eqref{eq:varXi_atom}.
For a small $E_\JJ^\cpl / E_L^\res$, we mention that $\phi_\res^\mi \sim E_\JJ^\cpl / E_L^\res$ according to Eq.~\eqref{eqResonatorExtremum}.
That allows us to obtain the leading terms in Eq.~\eqref{eq:varXi_res},
\begin{equation}
\label{eq:varXi_res_approx}
	\varXi_\res \approx \frac{E_C^\res}\hbar 
		\Bigg[\frac53 \Big(\frac{E_\JJ^\cpl}{E_L^\res}\Big)^2 \sin^2 \delta
			 + \frac{E_\JJ^\cpl}{E_L^\res} \cos\delta
		\Bigg],
\end{equation}
where we have also used Eq.~\eqref{eqRenormELr}.

In the Hamiltonian~\eqref{eqResonatorHamiltonian}, $V'_{\res3}$ and $V_{\res4}$ are the off-diagonal terms analogous to those in the atom Hamiltonian~\eqref{eqAtomHamiltonian}.
They are obtained by changing the atom notations to the resonator ones in Eqs.~\eqref{eqOffDiagonal3} and \eqref{eqOffDiagonalBare4}.

To arrive at the full system Hamiltonian, one also applies the above Schrieffer-Wolff transformations to the coupling part of the circuit Hamiltonian in Appendix~\ref{apSecondQuantizedHamiltonian}.
After rearrangement, we write out the full Hamiltonian
\begin{align} \label{eqSecondQuantizedHamiltonian3}
	\frac{\opH}{\hbar} \approx {}& \omega_\res a^\dag a - \frac{\varXi_\res}{2}a^{\dag 2} a^2 + V'_{\atom3} + V_{\atom4}
	\nonumber
	\\
	& + \omega_\atom b^\dag b - \frac{\varXi_\atom}{2} b^{\dag 2} b^2 + V'_{\res3} + V_{\res4} + \frac{\opH_\mathrm{I}}{\hbar}
\end{align}
up to the zero-point energy.
Here, $\omega_\res = \omega^p_\res - \varXi_\res - K_{0,X}/2$ and $\omega_\atom = \omega^p_\atom - \varXi_\atom - K_{0,X}/2$ denote the resonator and the atom frequencies with
\begin{equation} \label{eqCrossKerrMainFull}
	K_{0,X} \approx K_0 + 24 \Lambda_\atom g_2 + 24\Lambda_\res G_2.
\end{equation}
We obtain Eq.~\eqref{eqCrossKerrMainPartHalfBaked} in Sec.~\ref{secCrossKerr} by using Eq.~\eqref{eqLambdaAtomDefinition} and neglecting the last term in the above expression.
That is always possible when the resonator cubic nonlinearity is much weaker than in the atom, i.e., when $|\Lambda_\res| \ll |\Lambda_\atom|$;
and when, at the same time, $|g_2| \sim |G_2|$.
The latter is true when the vacuum fluctuations in the resonator and the artificial atom $\varphi_{\res,\zpf} \sim \varphi_{\atom,\zpf}$ are comparable.
A more general criterion for the approximation in Eq.~\eqref{eqCrossKerrMainPartHalfBaked} to hold, $(E_L^\atom / E_\JJ^\atom + \cot\phi^\mi_\atom) \sin\delta \lesssim 1$, is obtained by taking into account the condition~\eqref{eqSizableEJcNoUltrastrong} of no ultrastrong coupling.
Using the definition of $\Lambda_\atom$ in Eq.~\eqref{eqLambdaAtomDefinition} along with the expressions for $X_\atom$ and $g_2$ from Table~\ref{tablBareInteractions} in the above expression for $K_{0,X}$, one obtains Eq.~\eqref{eqCrossKerrMainPart} in Sec.~\ref{secCrossKerr}.

In Eq.~\eqref{eqSecondQuantizedHamiltonian3}, the Hamiltonian $\opH_\mathrm{I}$ aggregates all interaction terms;
as it is cumbersome, we do not present its full form.
However, by proper adjustment of the resonator and the atom frequencies, one can make their two-photon coupling dominant.
Most of other couplings then act as perturbations and renormalize the resonant interactions.
We treat them in what follows.

\subsection{Two-photon regime} \label{apTwoPhotonDerivation}

With $\omega_\atom \approx 2 \omega_\res$, close to the two-photon resonance, the interaction Hamiltonian becomes
\begin{equation} \label{eq:ham_int2}
	\frac{\opH_\mathrm{I}}{\hbar} \approx -g_2 (a^{\dag 2} b + b^\dag a^2) - K_{0,X} a^\dag a \, b^\dag b + \frac{\delta\opH_\mathrm{I}}{\hbar}.
\end{equation}
The first two terms here correspond to the resonant interactions:
the bare two-photon coupling of strength $g_2$ and the cross-Kerr coupling of strength $K_{0,X}$ as given by Eq.~\eqref{eqCrossKerrMainFull}.
Below, we calculate the essential renormalizations to these couplings.

Other interaction terms in the Hamiltonian~\eqref{eq:ham_int2} are small or nonresonant.
We collect the nonresonant ones that produce nonnegligible renormalizations into
\begin{equation} \label{eqHamiltonianPert}
	\begin{split}
		\frac{\delta\opH_\mathrm{I}}{\hbar} \approx {}& {- \tilde g_-} (a^\dag b + b^\dag a) - \tilde g_+ (a^\dag b^\dag + b a)
		\\
		& - 3 g_3 (b^\dag a^\dag a^2 + a^{\dag 2} a b) - 3 g_3 (b^\dag a^{\dag 2} a + a^\dag a^2 b)
		\\
		& - F (a^\dag b^\dag b^2 + b^{\dag 2} b a) - F (a^\dag b^{\dag 2} b + b^\dag b^2 a)
		\\
		& - g_2 (2 a^\dag a + 1) (b^\dag + b) - J (2b^\dag b + 1) (a^\dag + a)
		\\
		& - g_2 (a^{\dag 2} b^\dag + a^2 b)
		\\
		& - \tilde G^-_2 (a^\dag b^2 + b^{\dag 2} a) - \tilde G^+_2 (a^\dag b^{\dag 2} + b^2 a).
	\end{split}
\end{equation}
We treat $\delta\opH_\mathrm{I}$ as a perturbation in Sec.~\ref{apEliminateLinAndOptMech}.
Below we describe its terms.
Than, in Secs.~\ref{apSaApproximations} and \ref{apCouplingApproximations}, we discuss the approximations there.

The first two terms in the perturbation Hamiltonian~\eqref{eqHamiltonianPert} correspond to the number-conserving and number-nonconserving parts of the linear interaction as described in Table~\ref{tablBareInteractions}.
Their coupling rates $\tilde g_\pm = g_1^c \pm \tilde g_1^i$ include contributions from both capacitive and inductive couplings.
The latter coupling strength renormalizes as
\begin{equation}
\label{eqInductiveCouplingRenormInAppendix}
	\tilde g^i_1 \approx g^i_1 + 20 \Lambda_\atom G_2,
\end{equation}
where we neglect a contribution of $3 (g_3 + G_3)$ from the fourth-order expansion term in the coupler Josephson energy.
Indeed, when $E^\res_C \lll \tilde{E}^\res_L$ and $E^\atom_C \lll \tilde{E}^\atom_L$, the relevant vacuum fluctuations are small with $\varphi^2_{\res, \zpf}, \varphi^2_{\atom, \zpf} \ll 1$.
Thus, $3 |g_3 + G_3| \ll |g^i_1|$ as follows from the definition of $g^i_1$ in Table~\ref{tablBareInteractions} and $g_3$ with $G_3$ in Appendix~\ref{apSecondQuantizedHamiltonian}.
As for the second term in Eq.~\eqref{eqInductiveCouplingRenormInAppendix}, it matters near the points $\delta \mod \pi = \pi/2$ where the bare inductive coupling $g^i_1$ vanishes.

Away from there, the entire renormalization of the linear coupling $g_1^i$ often can be negligible.
The coupling strength $G_2$ is of higher order in the vacuum fluctuations compared to $g_1^i$.
Thus, we have that $20\Lambda_\atom |G_2| \ll |g^i_1|$ as far as $\mu \varphi^2_{\atom,\zpf} |\tan\delta| \ll 1$.
In that case, $\tilde g^i_1 \approx g^i_1$ according to Eq.~\eqref{eqInductiveCouplingRenormInAppendix}.

The second and third lines in Eq.~\eqref{eqHamiltonianPert} describe nonlinear interactions that result in the same transitions as the linear coupling.
The number-conserving part of such interactions is known as a nonlinear Jaynes-Cummings interaction (see Ref.~\cite{torres2025perfectrevivals} and the references therein).
Unlike the regular linear interactions, excitation and de-excitation of the atom occur in two steps here.
Consider the second line in Eq.~\eqref{eqHamiltonianPert} describing couplings of rate $3g_3$, whose energies are cubic in the resonator variables.
Here, during the excitation, the resonator first goes two energy levels up and then one level down.
Up to the order of creation and annihilation processes, this interaction is a linear coupling with its strength modulated by the photon number $a^\dag a$.
The third line in Eq.~\eqref{eqHamiltonianPert} contains analogous terms with the roles of the atom and resonator reversed.
However, the rate of these couplings renormalizes as
\begin{equation} \label{eqCouplingF}
	F = 3G_3 + 20\Lambda_\atom G_2,
\end{equation}
similarly to the linear inductive coupling in Eq.~\eqref{eqInductiveCouplingRenormInAppendix} and as described in Sec.~\ref{secDipolarInteractions} of the main text.

The fourth line in Eq.~\eqref{eqHamiltonianPert} describes the couplings of optomechanical type.
One of them has its strength renormalized as
\begin{equation} \label{eq:J1_2}
	J = G_2 + 6\Lambda_\atom g^i_1.
\end{equation}
Plugging $g^i_1$ and $G_2$ from Table~\ref{tablBareInteractions} along with Eqs.~\eqref{eqLambdaAtomDefinition} and \eqref{eqLambdaInTermsOfZPF} for $\Lambda_\atom$ yields the expression in Appendix~\ref{apOptomechanical}.

The penultimate line in Eq.~\eqref{eqHamiltonianPert} describes the energy-nonconserving two-photon coupling.
With rate of $g_2$, it simultaneously excites or de-excites the resonator by one energy level and the atom by two energy levels.
The last line in Eq.~\eqref{eqHamiltonianPert} describes two--atom-excitation interactions illustrated in Table~\ref{tablBareInteractions}.
They are dressed as
\begin{equation} \label{eqRenormalizedG2}
	\tilde G^\pm_2 = G_2 - 2\Lambda_\atom (g^i_1 \pm 2g^c_1).
\end{equation}
The second term arises from the capacitive and inductive couplings as the atom 
dresses by its cubic nonlinearity.

\subsubsection{Further approximations in treating self-nonlinearities}
\label{apSaApproximations}

In Sec.~\ref{apSelfNonlinearitiesTreatment}, we restrict the perturbation theory to second order in the cubic perturbation.
We need further simplifications to arrive at Hamiltonian in Eqs.~\eqref{eq:ham_int2} and \eqref{eqHamiltonianPert}.

First, consider the off-diagonal non-coupling terms in Eqs.~\eqref{eqOffDiagonal3} and \eqref{eqOffDiagonalBare4}.
Recall that $V'_{\atom3}$ arises on transforming the atom bare Hamiltonian in Eq.~\eqref{eqHaBare} while $V_{\atom4}$ is the off-diagonal part of its quartic nonlinearity.
The coefficients in $V'_{\atom3}$ and $V_{\atom4}$ are roughly as the atom anharmonicity $\varXi_\atom$ in the magnitude.
In the second-order perturbation theory, that brings different energy terms of about $\hbar\varXi_\atom^2 / \omega_\atom$.
They are small for small anharmonicity
\begin{equation}
\label{eqSmallAtomAnharmonicity}
	\varXi_\atom \ll \omega_\atom.
\end{equation}
We provide specific criteria to neglect such terms in the frequencies, as well as in the single-photon and two-photon coupling rates.
Note that $V'_{\res3}$ and $V_{\res4}$ give rise to even smaller contributions as the induced resonator anharmonicity is small with $|\varXi_\res| \ll |\varXi_\atom|$.

Similarly to the cubic terms in Sec.~\ref{apSelfNonlinearitiesTreatment}, one can approximately eliminate $V'_{\atom3}$ and $V_{\atom4}$ with a Schrieffer-Wolff transformation $\exp[(\frac{3\Lambda_\atom X_\atom - Y_\atom}{4\omega_\atom} b^{\dag 4} + \frac{12\Lambda_\atom X_\atom + 4Y_\atom}{2\omega_\atom} b^{\dag 3} b + \frac{18\Lambda_\atom X_\atom + 6Y_\atom}{2\omega_\atom} b^{\dag 2} + \frac{8\Lambda_\atom Y_\atom}{5\omega_\atom}b^{\dag 5} - \frac{24 \Lambda_\atom Y_\atom}{3\omega_\atom} b^{\dag 4} b - \frac{112\Lambda_\atom Y_\atom}{\omega_\atom} b^{\dag 3} b^2 - \frac{48\Lambda_\atom Y_\atom}{3\omega_\atom} b^{\dag 3} - \frac{336\Lambda_\atom Y_\atom}{\omega_\atom} b^{\dag 2} b - \frac{136\Lambda_\atom Y_\atom}{\omega_\atom} b^\dag) - \mathrm{h.c.}]$.
Applying it to the Hamiltonian~\eqref{eqAtomHamiltonian} renormalizes the atom frequency by about $-(\omega_\atom/12) (\mu^2 + \eta)^2 \varphi^4_{\atom,\zpf}$.
Under normal conditions, this dressing is smaller than the two-photon coupling strength and hence does not affect the resonance.
Roughly, that is the case for $|g_2| > \varXi_\atom^2 / \omega_\atom$.
Note that, upon transformation, the atom anharmonicity also acquires a dressing;
it is small when Eq.~\eqref{eqSmallAtomAnharmonicity} holds.

Elimination of the off-diagonal terms in Eq.~\eqref{eqAtomHamiltonian} dresses the interactions in Eq.~\eqref{eqHamiltonianPert} as well.
Particularly, that adds $(\mu^2 + \eta) \varphi^2_{\atom,\zpf} g^i_1/8$ into the inductive coupling rate $g_i$.
However, this correction is smaller than $3(g_3 + G_3)$ neglected in Eq.~\eqref{eqInductiveCouplingRenormInAppendix} since $(\mu^2 + \eta)/8 < 1$ for $\mu \lesssim 2$.
The two-photon coupling also gets dressed by the terms in Eq.~\eqref{eqOffDiagonal3}.
It acquires a correction of $(\mu^2 + \eta)\varphi^2_{\atom, \zpf} g_2/8$.
However, it is of the fifth order in the zero-point fluctuations and of the first order in the coupler energy $E^\cpl_\JJ$. 
Hence it exceeds the accuracy of our approximation in the Hamiltonian~\eqref{eqHcBare} regardless of $E^\cpl_\JJ$ magnitude.
From the other point of view, one mentions that both corrections are small with  a relatively small atom anharmonicity as in Eq.~\eqref{eqSmallAtomAnharmonicity}.

\subsubsection{Approximations in coupling dressed by the cubic perturbation}
\label{apCouplingApproximations}

In Hamiltonian~\eqref{eq:ham_int2}, we group the ``new'' perturbations into $\delta\opH_\mathrm{I}$ in Eq.~\eqref{eqHamiltonianPert}.
There, we retain only the terms that provide essential renormalizations.
Dressing the coupling terms in the bare Hamiltonian~\eqref{eqHcBare} of Appendix~\ref{apSecondQuantizedHamiltonian} not only renormalizes them:
That also creates new couplings different from those in the bare interaction Hamiltonian~\eqref{eqHcBare}.
However, they are of magnitudes comparable to the same-type terms that we neglect in the cosine expansion~\cite{parti} of the coupler energy in Eq.~\eqref{eqHcBare}.
In other words, they are of the same orders in the zero-point fluctuations $\varphi_{\res,\zpf}$ and $\varphi_{\atom,\zpf}$ and the Josephson energy of the coupler $E_\JJ^\cpl$.
Therefore, we omit these new couplings in Eq.~\eqref{eqHamiltonianPert} as they exceed the accuracy of our approximation in the Hamiltonian~\eqref{eqHcBare}.

In addition to the negligible new couplings, small coupling terms of the same type as in the Hamiltonian~\eqref{eqHcBare} also arise in Eq.~\eqref{eq:ham_int2}.
They read as
\begin{equation} \label{eqDroppedTerms2}
\begin{split}
  	\frac{\delta \hat h_\mathrm{I}}{\hbar} =
		{}& {-\tilde G_3} (a^\dag + a)(b^{\dag 3} + b^3)
\\
		{}& - g_3 (a^{\dag 3} + a^3)(b^\dag + b)
\\
		{}& - \frac{\tilde K_0}{4}(a^{\dag} + a)^2 (b^{\dag 2} + b^2)
\\
		{}& - \frac{K_{0,X}}{4} (2b^\dag b + 1)(a^{\dag 2} + a^2)
\end{split}
\end{equation}
with $\tilde{G}_3 = G_3 - 4\Lambda_\atom G_2$ and $\tilde{K}_0 = K_0 - 8\Lambda_\atom g_2$.
These terms yield only negligible corrections to the frequencies and to the cross-Kerr and two-photon interaction rates.
We have verified that such corrections appear at least in the seventh order in the zero-point fluctuations $\varphi_{\res,\zpf}$ and $\varphi_{\atom,\zpf}$.
Hence they are beyond the accuracy of our expressions.

In Eqs.~\eqref{eqHamiltonianPert} and \eqref{eqDroppedTerms2}, we account for the corrections proportional to $\Lambda_\atom$ in the coupling rates.
At the same time, we neglect the other corrections with $\Lambda_\res$, since $|\Lambda_\res| \ll |\Lambda_\atom|$.

\subsubsection{Elimination of nonresonant linear and optomechanical interactions}
\label{apEliminateLinAndOptMech}

Here we calculate corrections in the two-photon coupling and the effective resonator anharmonicity.

The perturbation $\delta\opH_\mathrm{I}$ can be approximately eliminated via the Schrieffer-Wolff transformation $\opH \rightarrow e^{-\opW} \opH e^{\opW}$, where the anti-Hermitian operator $\opW$ reads as
\begin{equation} \label{eqAntiHermitianOperator}
	\begin{split}
		\opW = {} & \lambda_{-} (a^\dag b - b^\dag a) + \lambda_{+} (a^\dag b^\dag - b a)
		\\
		& {} + \zeta_- (b^{\dag 2} b a - a^\dag b^\dag b^2) + \zeta_+ (a^\dag b^{\dag 2} b - b^\dag b^2 a)
		\\
		& {} + \vartheta_- (a^{\dag 2} a b - b^\dag a^\dag a^2) + \vartheta_+ (b^\dag a^{\dag 2} a - a^\dag a^2 b)
		\\
		& {} + \lambda_\res \left(a^\dag a + \tfrac{1}{2}\right)(b - b^\dag) + \lambda_\atom \left(b^\dag b + \tfrac{1}{2}\right) (a - a^\dag)
		\\
		& {} + \lambda_{2}(a^{\dag2} b^\dag - b a^2)
		\\
		& {} + \eta_+(a^\dag b^{\dag 2} - b^2 a) + \eta_- (b^{\dag 2} a - a^\dag b^2).
	\end{split}
\end{equation}
The small parameters $\lambda_\pm,\lambda_\res,\lambda_\atom, \lambda_2, \zeta_\pm, \vartheta_\pm, \eta_\pm \ll 1$ read
\begin{equation} \label{eqSmallParameters}
	\begin{split}
		& \lambda_\pm = \frac{\tilde g_\pm}{\omega_\atom \pm \omega_\res},
		\quad \zeta_\pm = \frac{F}{\omega_\atom \pm \omega_\res - \varXi_\atom},
		\quad \vartheta_\pm = \frac{3g_3}{\omega_\atom \pm \omega_\res},
		\\
		& \lambda_\res = -\frac{2g_2}{\omega_\atom}, \quad \lambda_\atom = -\frac{2 J}{\omega_\res}, \quad \lambda_{2} = \frac{g_2}{\omega_\atom + 2\omega_\res},
		\\
		& \eta_\pm = \frac{\tilde{G}^\pm_2}{2\omega_\atom \pm \omega_\res - \varXi_\atom}.
	\end{split}
\end{equation}
Now we use the approximate BCH formula as for Eqs.~\eqref{eqAtomHamiltonian} and \eqref{eqResonatorHamiltonian}.
With the operator $\opW$ and the Hamiltonian in Eqs.~\eqref{eqSecondQuantizedHamiltonian3}, \eqref{eq:ham_int2}, and \eqref{eqHamiltonianPert}, that yields
\begin{equation}
\label{eqSecondQuantizedHamiltonian4}
\begin{split}
	\frac{\opH}{\hbar} \approx {}& \omega_\res a^\dag a
		- \frac{\varXi'_\res}{2}a^{\dag 2} a^2
		+ \omega_\atom b^\dag b - \frac{\varXi'_\atom}{2} b^{\dag 2} b^2
\\
	& {} -K_{0,X} a^\dag a b^\dag b - \tilde g_2 (a^{\dag 2} b + b^\dag a^2).
\end{split}
\end{equation}
Here, we have omitted the higher-order nonlinearities exceeding the accuracy of the approximation in Eqs.~\eqref{eqHrBare} and \eqref{eqHaBare}.

In this contribution, we omit further dressing of the transition frequencies $\omega_\res$ and $\omega_\atom$.
They are of the second order in the small parameters in Eqs.~\eqref{eqSmallParameters}.
We assume that the two-photon coupling rate is much larger than these corrections and hence they do not influence the resonance condition.
We also omit the anharmonicity dressing, as well as the further dressing in the cross-Kerr coupling as discussed in Sec.~\ref{secCrossKerr}.
Note that calculation of the cross-Kerr dressing by other interactions requires taking into account $V'_{\atom3}$ and $V_{\atom4}$ in Eqs.~\eqref{eqOffDiagonal3} and \eqref{eqOffDiagonalBare4}.
We leave derivation and analysis of these corrections for future work.

Here we only calculate the renormalized two-photon coupling strength,
\begin{equation} \label{eq:g2_explicit_renorm}
	\begin{split}
	 \tilde g_2 = {} & g_2 + \frac{\lambda_\atom}{2} g_- - \lambda_{-} J
	 \\ & + \eta_- g_+ + \lambda_+ \tilde{G}^-_2 + 2 \lambda_+ \eta_- \varXi_\atom.
	\end{split}
\end{equation}
We have dropped the corrections $\zeta_+ \tilde{G}^-_2 + \eta_- F + \tfrac{3}{2}(\vartheta_- J - \tfrac{3}{2}\lambda_\atom g_3) + 2\eta_-\zeta_+\varXi_\atom$ as they are of the seventh order in the zero-point fluctuations $\varphi_{\res,\zpf}$ and $\varphi_{\atom,\zpf}$ and hence beyond the accuracy of the Hamiltonian in Eq.~\eqref{eqHcBare}.
We leave the terms of the fifth order in $\varphi_{\res,\zpf}$ and $\varphi_{\atom,\zpf}$.
They can be relevant as explained in Sec.~\ref{secTwoPhotonIsBare}.

\section{Calculation of the main cross-Kerr correction with Wick's theorem}
\label{apWick}

On the transformations in Appendix~\ref{apRenormHamiltonianDerivation}, the interactions we are interested in pick up corrections.
Each transformation is designed to compensate the direct action of a nonresonant coupling in the resulting frame.
With a nonlinear nonresonant coupling, that requires a polynomial of the same order in the transformation exponent.
When a transformation is complicated, such as in Eq.~\eqref{eqAtomTransformation}, it becomes nontrivial to trace the origin of the interaction renormalizations.

That is especially concerning for a phase qubit as in Sec.~\ref{secEstimates}.
Its metastable well holds a finite number of energy states;
hence a virtual process climbing too high does not occur as predicted by our anharmonic oscillator model.
It is then of interest to determine how far up the processes in each correction do climb.

In this appendix, we demonstrate how to use Wick's theorem~\cite{wick1950evaluation,blasiak2007combinatorics} to calculate the coupling corrections.
That provides some grounding for laying out the virtual processes in the diagrams in Figs.~\ref{figDiagramsLinearAndCrossKerr} and \ref{figDiagramsTwoPhoton}.
For the phase qubit as in Sec.~\ref{secEstimates}, the most vivid correction due to the virtual processes appears in the cross-Kerr coupling strength.
We, therefore, focus on the cross-Kerr correction here.

\subsection{Calculation}

Beforehand, it is convenient to rewrite Eq.~\eqref{eqAtomTransformation} that we use in Appendix~\ref{apRenormHamiltonianDerivation} as
\begin{equation}
\label{eqSimpleTransformation}
	\opS_\atom = -\frac23(b^\dag - b)^3
		+ (b^\dag + b)^2 (b^\dag - b) - 2(b^\dag + b).
\end{equation}
We verify this form with \textsc{pyBoLaNO}~\cite{lim2025pybolano} which is based on Refs.~\cite{blasiak2005thesis,mendez2006combinatorial}.
With Eq.~\eqref{eqSimpleTransformation}, the commutators in the BCH formula~\eqref{eqBCH} are easy to calculate, as our bare coupling Hamiltonian~\eqref{eqHcBare} in Appendix~\ref{apSecondQuantizedHamiltonian} depends on the dimensionless canonical coordinates $b^\dag + b$ and $i(b^\dag - b)$.
Note that Refs.~\cite{zueco2009qubit,petrovnin2023microwave} provide their starting Hamiltonians in such form.

Below, we show that dressing of the coupling energy
\begin{equation}
	I_{20} = \hbar g_2 (a + a^\dag)^2 (b + b^\dag)
\end{equation}
quadratic in the resonator operators [see Eq.~\eqref{eqHcBare}] gives rise to the cross-Kerr coupling.
More precisely, the cross-Kerr correction in Eqs.~\eqref{eqCrossKerrMainPartHalfBaked} and \eqref{eqCrossKerrMainPart} arises from expanding the transformation exponent to the first order in $\Lambda_\atom$,
\begin{equation}
\label{eqCrossKerrFirstCommutator}
	I_{20} \to e^{-\Lambda_\atom \opS_\atom} I_{20} e^{\Lambda_\atom \opS_\atom} \approx I_{20} + \Lambda_\atom (I_{20} \opS_\atom - \opS_\atom I_{20}).
\end{equation}
Here $\Lambda_\atom = X_\atom / \omega^p_\atom$ as defined in Eq.~\eqref{eqLambdaAtomDefinition}.

While it is trivial to calculate our commutator in Eq.~\eqref{eqCrossKerrFirstCommutator}, the calculation can become more involved for higher nonlinearities.
We use Wick's theorem as it makes the manual treatment tractable for more complex cases.
Additionally, it yields the result in normal order.

Thus, we use it to calculate the correction, i.e., the last term in the right-hand side of Eq.~\eqref{eqCrossKerrFirstCommutator}.
We denote
\begin{equation}
	P = b^\dag + b,
\quad
	M = b^\dag - b.
\end{equation}
Then, noting that only single contractions bring the terms with $b^\dag b$, we write them out for
\begin{multline}
\label{eqPS}
	(b^\dag + b) \opS = -\frac{2\times3}3 \wick{: \c P \c M M M :} - \frac23 C_3^2 \wick{: P \c M M \c M:}
\\
	+ 3 \wick{: P P \c P \c M :} - C_3^2 : \wick{\c P \c P P M} : + \ldots
\end{multline}
and
\begin{equation}
\label{eqSP}
	\opS (b^\dag + b) = 2 :\wick{P \c P \c M P} : - :\wick{P \c P \c M P} : + \ldots,
\end{equation}
where $C_n^k = n! / [k!(n-k)!]$ denotes a binomial coefficient, $:O:$ denotes the operation of normal ordering applied to an operator $O$, and contraction is defined as $\wick{\c A \c B} = AB - :AB:$.

In Eq.~\eqref{eqPS}, the first two terms are the contractions with the first term $-2M^3/3$ in Eq.~\eqref{eqSimpleTransformation}.
The other two are the contractions with the second term $P^2 M$ there.
In Eq.~\eqref{eqSP}, both terms are due to the second term $P^2 M$ in Eq.~\eqref{eqSimpleTransformation}.
Each contraction coefficient includes a factor that counts the contractions alike, such as $\wick{: \c P \c M M M :}$, $\wick{: \c P M \c M M :}$, and $\wick{: \c P M M \c M :}$ for the first term in Eq.~\eqref{eqPS}.

We note that $\wick{\c P \c P} = 1$, $\wick{\c M \c M} = -1$, and $\wick{\c P \c M} = 1$.
Then using Eqs.~\eqref{eqPS} and \eqref{eqSP}, one can check that the cross-Kerr term proportional to $a^\dag a \, b^\dag b$ comes with a coefficient of $24 g_2 \Lambda_\atom$ in Eq.~\eqref{eqCrossKerrFirstCommutator}.
That is the same coefficient as in Eqs.~\eqref{eqCrossKerrMainPartHalfBaked} and \eqref{eqCrossKerrMainFull} according to the calculation of the correction in Appendix~\ref{apRenormHamiltonianDerivation}.

\subsection{Diagrams and the number of energy levels}

The atom part of the diagram in Fig.~\ref{figDiagramsLinearAndCrossKerr}(I) shows the process as described, among others, by the term $\wick{: P \c M \c M M:}$ similar to the second term in Eq.~\eqref{eqPS}.
A contraction in a term corresponds to a loop in the diagram.
Note that, when only taking into account the first BCH commutator as in Eq.~\eqref{eqCrossKerrFirstCommutator}, the processes from different interactions do not mix.

The normal-ordered products as in Eqs.~\eqref{eqPS} and \eqref{eqSP} directly provide the cross-Kerr coupling Hamiltonian in a convenient form.
Wick's theorem automatically takes into account that the $bb^\dag$ products in Eq.~\eqref{eqCrossKerrFirstCommutator} yield the normally-ordered $b^\dag b$.
The associated use of commutation relations---bosonic in our case---is exact within the model;
however, with it we loose the information on how far up in the energy ladder a process goes.

Based on the structure of the normally-ordered terms in Eqs.~\eqref{eqPS} and \eqref{eqSP}, we can recover that from Eq.~\eqref{eqCrossKerrFirstCommutator}.
In the case of the cross-Kerr correction, we simply notice that some of the $b^\dag b$ processes actually arise from the $bb^\dag$ ones that bring the atom higher in its energy levels.
One such sequence is depicted in Fig.~\ref{figDiagramsLinearAndCrossKerr}(I) with dashed lines.

Note that, for determining the number of energy states required, one can consider normally-ordered $I_{20}$ and $\opS_\atom$ in Eq.~\eqref{eqCrossKerrFirstCommutator}.
In this form, each of them describe processes that excite the atom as low as possible while still providing the required results.
For example, consider the case when the next higher energy state decays faster than the interaction can occur, due to the reasons \emph{beyond} our simple anharmonic oscillator model.
In that case, the coupling energy $I_{20}$ cast in the normally-ordered form, clearly describes the interactions in the same way.
Same considers $\opS_\atom$ which is responsible in Eq.~\eqref{eqCrossKerrFirstCommutator} for an approximate unitary transformation.

That is not the case for further normal ordering in the equation.
In Eq.~\eqref{eqCrossKerrFirstCommutator}, $bb^\dag$ and $b^\dag b + 1$ can differ as the bosonic approximation breaks.
If the excitations in $\opS_\atom$ or $I_{20}$ bring the atom into, e.g., a rapidly decaying state, the respective processes yield different states.
Our approximate unitary transformation does not work as intended, as the corresponding (virtual) processes occur differently.
However, these very processes are supposed to yield the interactions in the transformed picture.
Therefore, if we further normal-order Eq.~\eqref{eqCrossKerrFirstCommutator}, this can hide such an issue and may predict interaction terms in the Hamiltonian with remarkably incorrect amplitudes or even structure.

\section{Highest-climbing processes}
\label{apRules}

In Figs.~\ref{figDiagramsLinearAndCrossKerr} and \ref{figDiagramsTwoPhoton}, the correction processes climb the atom energy ladder as high as possible.
We provide a recipe to determine such processes:
\begin{enumerate}

\item
Determine a Schrieffer-Wolff transformation to take into account non-resonant interactions perturbatively.
For several transformations, the recipe applies recursively.
For example, see Fig.~\ref{figDiagramsTwoPhoton}(b) explained in Appendix~\ref{apTwoPhotonFormula}.

\item
\label{itemFormInteraction}
Take one operator sequence from an interaction term in the Hamiltonian and another one from the transformation exponent.
It is convenient to assign a color for each sequence.
The available operators should accomplish the required interaction.
One can vary the order of sequences but cannot intermix operators of different color.
One may contract pairs of creation and annihilation operators into unity.

\item
\label{itemContractions}
Show a loop heading upwards when contracting $bb^\dag$ and a loop downwards when contracting $b^\dag b$.

\item
\label{itemSingleColorLoops}
Only two-color loops require next energy level to be available.
Hence no energy levels on top of the single-color loops in Figs.~\ref{figDiagramsLinearAndCrossKerr} and \ref{figDiagramsTwoPhoton}.

\item
Consider a sequence that increases the energy more than the other ones, possibly in the midst of the sequence.
Then place it next to the highest of the two energy levels at the ends of the interaction to be realized.

\end{enumerate}

Let us comment on some of the rules.
Consider rule~\ref{itemFormInteraction}.
The sequence order can be changed as formally the whole of each correction stems from a commutator (see Appendix~\ref{apWick}), where the sequences occur in both orders.
Contractions in rules~\ref{itemFormInteraction} and~\ref{itemContractions} are discussed in Appendix~\ref{apWick} and in the caption of Fig.~\ref{figDiagramsLinearAndCrossKerr}.
For motivation of rule~\ref{itemSingleColorLoops}, refer to Appendix~\ref{apWick}.

\section{Calculation of the main cross-Kerr correction using dynamical equations}
\label{apDynamicalEqs}

Consider, for simplicity, the Hamiltonian $H/\hbar = \omega_\res a^\dag a + \omega_\atom b^\dag b - g_2 (a^\dag + a)^2 (b^\dag + b) - X_\atom (b^\dag + b)^3.$
It contains the necessary terms to obtain the cross-Kerr correction due to the potential asymmetry and nonlinear coupling.
Here we obtain this cross-Kerr rate by analyzing the dynamical equations and their formal solutions, following from the Hamiltonian.

The Heisenberg equation of motion for the atom lowering operator is obtained as $\dot b = \frac1{i\hbar}[b, H]$ and reads
\begin{equation}
\label{eqDiffb}
	\dot b = -i\omega_\atom b + 3iX_\atom (b^\dag + b)^2 + ig_2 (a^\dag + a)^2.
\end{equation}
A formal solution of this equation is
\begin{multline}
\label{eqbFormalSolution}
    b(t) = b(0)e^{-i\omega_\atom t}
		+ 3iX_\atom \int_0^t dt' e^{-i\omega_\atom (t-t')} (b^\dag + b)^2_{t'}
\\
		+ ig_2 \int_0^t dt' e^{-i\omega_\atom (t-t')} (a^\dag + a)^2_{t'}.
\end{multline}
We assume that $a^\dag a$ changes slowly at the timescales of $t$ the observation time.
Integrating the last term yields $(g_2 / \omega_\atom) (2a^\dag a$ + 1) and some oscillating terms.
We plug that back into the dynamical Eq.~\eqref{eqDiffb} to obtain
\begin{align}
\label{eqDiffbWithCorrection}
	\dot b \approx &{-i}\bigg(\omega_\atom
						   - \frac{12 g_2 X_\atom}{\omega_\atom}\bigg) b
			+ \frac{24i g_2 X_\atom}{\omega_\atom} ba^\dag a
			+ ig_2 (a^\dag + a)^2
\nonumber
\\
	&{}+ \text{anharmonicity terms.}
\end{align}
Here we neglect the nonresonant contribution in the second term in the right-hand side.
The resulting term is approximately as obtained from the Hamiltonian with the cross-Kerr correction in Eq.~\eqref{eqCrossKerrMainPartHalfBaked} from Sec.~\ref{secCrossKerr}.

The atom frequency in Eq.~\eqref{eqDiffbWithCorrection} coincides with Eqs.~\eqref{eqFreqs} and \eqref{eqCrossKerrMainPartHalfBaked} with $K_0 = 0$.
Existence of the population-independent frequency shift $12 g_2 X_\atom / \omega_\atom$ relies on commuting $a$ and $a^\dag$ in Eq.~\eqref{eqbFormalSolution}.
Hence, unlike the cross-Kerr term, it vanishes in the classical limit.

\section{Other interaction dressing}
\label{apOtherInteractions}

In this appendix, we explain other small renormalizations in the resonant two-photon and in the nonresonant optomechanical-like couplings.
We often call the coupling types by their bare rates here.
For details, refer to Table~\ref{tablBareInteractions} in the main text and the bare coupling Hamiltonian~\eqref{eqHcBare} in Appendix~\ref{apSecondQuantizedHamiltonian}.

\subsection{Non-resonant optomechanical interactions}
\label{apOptomechanical}

As pointed out in Table~\ref{tablBareInteractions}, the coupling term with $G_2$ yields a ``reverse'' optomechanical-like interaction.
Owing to a combination of inductive interactions with the atom cubic nonlinearity, this optomechanical coupling renormalizes as $G_2 \to J$, with the full coupling rate
\begin{align}
\label{eqOptomechanicalCouplingAppendix}
	J &= G_2 + \frac{6 g^i_1 X_\atom}{\omega^p_\atom}
\\
\label{eqOptomechanicalCouplingMu}
	  &\approx \frac{E^\cpl_\JJ}{2\hbar}
				\big(\sin\delta - \mu \cos\delta\big)
				\, \varphi_{\res,\zpf} \varphi^2_{\atom,\zpf}.
\end{align}
This correction to the $G_2$ interaction can be interpreted similarly as in Eq.~\eqref{eqInductiveCouplingRenormHalfBaked} of the main text.

Within our assumptions, the optomechanical $g_2$ interaction retains its bare magnitude as in Table~\ref{tablBareInteractions}.
This asymmetry, as seen from Table~\ref{tablBareInteractions} and Eq.~\eqref{eqOptomechanicalCouplingMu}, is due to the cubic nonlinearity in the potential of the atom that is negligible in the resonator.

\subsection{Details on two-photon corrections}
\label{apTwoPhotonFormula}

Here we provide and interpret expressions for the two-photon corrections for reasonably strong Josephson coupling.
Assuming a relatively small atom anharmonicity, $|\varXi_\atom| \ll \omega_\atom$, we obtain from the results of Appendix~\ref{apTwoPhotonDerivation} that
\begin{equation}
\label{eqFullTwoPhotonCoupling}
	 \tilde g_2 \approx g_2 - \frac{2g_- J}{\omega_\res}
		+ \frac{2g_+ G_2}{3\omega_\res}
		- \frac{\mu g_+ (g^i_1 - 2g^c_1) \varphi_{\atom, \zpf}}
			   {9\omega_\res}
\end{equation}
for the two-photon coupling rate in the Hamiltonian~\eqref{eqTwoPhotonHamiltonian}.
The last three terms are due to various perturbative corrections.
As pointed out in Sec.~\ref{secTwoPhotonIsBare}, they are nonnegligible for higher coupling energies as in Eq.~\eqref{eqNonnegligibleCorrections}.
We discuss these corrections below.

Consider the correction in the second term in Eq.~\eqref{eqFullTwoPhotonCoupling}.
This correction is due to chaining of the single-photon and optomechanical virtual processes.
We depict one such sequence with a bare optomechanical process in Fig.~\ref{figDiagramsTwoPhoton}(a).
Another one arises through its dressed part in Eq.~\eqref{eqOptomechanicalCouplingAppendix}.
In other words, linear inductive interaction chains with the full number-conserving one and the atom cubic nonlinearity.
We depict one such sequence with a number-nonconserving part of the inductive interaction in Fig.~\ref{figDiagramsTwoPhoton}(b).

The correction as in the third term in Eq.~\eqref{eqFullTwoPhotonCoupling} can be depicted similarly to the second term, see Fig.~\ref{figDiagramsTwoPhoton}(c).
Here, however, a $G_2$ interaction that doubly (de-)excites the atom chains with the number-nonconserving linear interaction.
As before, in Fig.~\ref{figDiagramsTwoPhoton}(c) we choose the particular processes so that the other similar ones do not excite the atom higher.

Now consider the two-photon correction in the last term in Eq.~\eqref{eqFullTwoPhotonCoupling}.
It arises, among others, owing to a process similar to that in Fig.~\ref{figDiagramsTwoPhoton}(b) and uses four energy states of the atom.
This term is considerably smaller than the other corrections, owing to the coefficient of $\mu/9$.
Still, the term can be significant with even smaller zero-point fluctuations or when the capacitive coupling is much stronger than the inductive one $g_1^c \gg g_1^i$.

The latter case is unlikely when one aims at stronger two-photon coupling.
We expect $g_1^c$ to be of the same order or smaller than the maximal value of the inductive coupling rate $(g_1^i)_\text{max}$.
Both couplings are limited by the criterion $g_- = g_1^c - g_1^i \ll \omega_\res, \omega_\atom$ disallowing ultrastrong coupling.
Practically, one raises the coupler Josephson energy $E_\JJ^\cpl$ to reach higher two-photon coupling strength $g_2 \propto E_\JJ^\cpl/\hbar$;
that raises $g_1^i \propto E_\JJ^\cpl/\hbar$ proportionally.
Therefore, for such an application, it is unlikely that $g_1^c \gg (g_1^i)_\text{max}$, as that severely limits the magnitudes of $(g_1^i)_\text{max}$ and hence $g_2$.

In the derivation of Eq.~\eqref{eqFullTwoPhotonCoupling} in Appendix~\ref{apTwoPhotonDerivation}, we neglect perturbative corrections of seventh-order in the zero-point fluctuations $\varphi_{\atom,\zpf}$ and $\varphi_{\res,\zpf}$.
In principle, they can exceed the fifth-order terms in the expansion of the Josephson coupling energy.
However, that occurs with even higher coupling energies $E_\JJ^\cpl$ as in $\varphi^2_\zpf E_\JJ^\cpl \gtrsim \hbar\omega_\res$ assuming $\varphi_\zpf \sim \varphi_{\atom,\zpf} \sim \varphi_{\res,\zpf}$.
By the similar reasons, in Eq.~\eqref{eqFullTwoPhotonCoupling}, we use the bare value of the linear coupling rates $g_\pm$ as in Table~\ref{tablBareInteractions}.

\bibliography{bibliography,common_sources,counters,jj_sources,sokolov,cqed,nonlin,optmech,param,circulators,gates,comp,vacuum}

\end{document}

%% file: notations.tex
\newcommand{\ket}[1]{|{#1}\rangle} 

\newcommand{\Ham}{\mathcal{H}}			

\newcommand{\opH}{\hat\Ham}				
\newcommand{\opS}{\mathcal{S}}			
\newcommand{\opW}{\mathcal{W}}			

\newcommand{\res}{\mathrm{r}}  			
\newcommand{\atom}{\mathrm{a}}			
\newcommand{\cpl}{\mathrm{c}}			
\newcommand{\JJ}{\mathrm{J}}			
\newcommand{\ext}{\mathrm{e}}			
\newcommand{\mi}{\mathrm{min}}			
\newcommand{\zpf}{\mathrm{zpf}}			

\newcommand{\Phie}{\Phi_\ext}	
\newcommand{\Phic}{\Phi_\cpl}	

%% file: highlights.tex
\usepackage{color}

\definecolor{nblack}{rgb}{0.05, 0.25, 0.5}
\definecolor{nred}{rgb}{0.7, 0.2, 0.2}
\definecolor{nblue}{rgb}{0.2, 0.2, 0.75}
\definecolor{ngrey}{rgb}{0.25, 0.3, 0.5}


%% file: interactions-tabular.tex
\newcommand\topstrut{\rule{0pt}{1.3em}}
\newcommand\bottomstrut{\rule[-0.7em]{0pt}{0pt}}
\newcommand\fatbottomstrut{\rule[-1.4em]{0pt}{0pt}}

\begin{center}
\begin{ruledtabular}
\begin{tabular}{l c c c c}
&
	Energy
&
		Rate
&
			\multicolumn2c{Some processes}
\bottomstrut
\\
\hline
\topstrut
Potential asymmetry
&
	$-\hbar X_\atom (b^\dag + b)^3$
&
		$X_\atom = \frac{\mu}{12} \omega^p_\atom \varphi_{\atom,\zpf}$
&
			\multicolumn2c{%
			$X_\atom b b^\dag b + \text{h.~c.} = X_\atom$
			\includegraphics{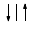}}
\bottomstrut
\\
\hline
\topstrut
Linear coupling
&
	capacitive
&&
			number-preserving
&
				-nonpreserving
\\
&	
	$-\hbar g_1^c (a^\dag - a)(b^\dag - b)$
&
		$g_1^c = \frac{E^\cpl_C}{\hbar} n_{\res,\zpf} n_{\atom,\zpf}$
&
			$g_- = g_1^c - g_1^i$
&
				$g_+ = g_1^c + g_1^i$
\bottomstrut
\\
\cline{2-3}
\topstrut
&
	inductive 
&&
		\multirow2*{$g_- a^\dag b + \text{h.~c.} = g_-$}
		\multirow2*{\includegraphics{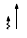}}
&
		\multirow2*{$g_+ ab + \text{h.~c.} =  g_+$}
		\multirow2*{\includegraphics{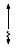}}
\\
&
	$-\hbar g_1^i (a^\dag + a)(b^\dag + b)$
&
		$g^i_1 = \frac{E^\cpl_\JJ}{\hbar}
		 \varphi_{\res,\zpf} \varphi_{\atom,\zpf} \cos\delta$
&
&
\bottomstrut
\\
\hline
\topstrut
Cubic coupling
&
	$-\hbar\frac{K_0}{4} (a^\dag + a)^2 (b^\dag + b)^2$
&
		$K_0 = \frac{E^\cpl_\JJ}{\hbar}
		 \varphi_{\res, \zpf}^2 \varphi_{\atom,\zpf}^2 \cos\delta$
&
			\multicolumn2c{cross-Kerr: $a^\dag a b^\dag b = {}$%
			\raisebox{-4pt}{%
			\includegraphics{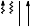}}}
\bottomstrut
\\
\hline
\topstrut
Quadratic couplings
&
	$-\hbar g_2 (a^\dag + a)^2 (b^\dag + b)$
&
		$g_2 = -\frac{E^\cpl_\JJ}{2\hbar}
		 \varphi^2_{\res, \zpf} \varphi_{\atom,\zpf} \sin\delta$
&
			\multicolumn2c{%
			twophoton:
			\multirow2*{%
			\includegraphics{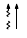}}%
			\multirow2*{$\quad$}
			``optomechanical'':
			\multirow2*{%
			\includegraphics{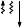}}
			} 
\\
\\
\cline{2-5}
\topstrut
&
	$-\hbar G_2 (a^\dag + a) (b^\dag + b)^2$
&
		$G_2 = \frac{E^\cpl_\JJ}{2\hbar}
		 \varphi_{\res,\zpf} \varphi^2_{\atom,\zpf} \sin\delta$
&
			\multirow3{12ex}{two--atom-\\excitation:\\}
			\multirow3*{
			\includegraphics{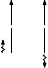}}
&
			\multirow3{19ex}{reverse\\``optomechanical'':\\}
			\multirow3*{
			\includegraphics{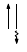}}
\\
\\
\\
\end{tabular}
\end{ruledtabular}
\end{center}